\documentclass{aa}
\usepackage{natbib}
\usepackage{multicol}
\usepackage{siunitx}
\bibpunct{(}{)}{;}{a}{}{,}
\usepackage{graphicx}
\usepackage{grffile}
\usepackage[varg]{txfonts}
\usepackage{pdfpages}
\usepackage{booktabs}
\usepackage{tabularx}
\usepackage{xspace}
\usepackage[normalem]{ulem}
\usepackage{xcolor}
\usepackage{hyperref}
\usepackage{float}

\defcitealias{lach2021a}{L21}

\newcommand{\msun}{\ensuremath{\mathrm{M}_\odot} }
\newcommand{\mni}{\ensuremath{M(^{56}\mathrm{Ni})} }
\newcommand{\nifs}{\ensuremath{^{56}\mathrm{Ni}} }
\newcommand{\mch}{\ensuremath{M_\mathrm{Ch}} }

\def\apjl{ApJ}%
\def\apjs{ApJS}%
\def\ao{Appl.~Opt.}%
\def\aap{A\&A}%
\titlerunning{Models of pulsationally assisted gravitationally confined
detonations}
\authorrunning{F. Lach et al.}

\begin{document}

\title{Models of pulsationally assisted gravitationally confined
detonations with different ignition conditions}

\author{F.~Lach\inst{1,2}\thanks{E-mail: florian.lach@h-its.org}
\and F.~P.~Callan\inst{3}
\and S.~A.~Sim\inst{3}
\and F.~K.~Roepke\inst{1,2}
}
  
\institute{ Heidelberger Institut f\"{u}r Theoretische Studien, 
            Schloss-Wolfsbrunnenweg 35, D-69118 Heidelberg, Germany\\
  \and Zentrum f{\"u}r Astronomie der Universit{\"a}t Heidelberg, 
       Institut f{\"u}r Theoretische Astrophysik, Philosophenweg 12, 
       D-69120 Heidelberg, Germany\\
  \and School of Mathematics and Physics, Queen’s University
       Belfast, Belfast BT7 1NN, UK \\
}

\date{15 November 2021 / 24 November 2021}

 \abstract{
 Over the past decades, many explosion scenarios for Type Ia supernovae
 have been proposed and investigated including various combinations of
 deflagrations and detonations in white dwarfs of different masses up to
 the Chandrasekhar mass. One of these is the gravitationally confined
 detonation model. In this case a weak deflagration burns to the
 surface, wraps around the bound core, and collides at the antipode. A
 subsequent detonation is then initiated in the collision area.  Since the
 parameter space for this scenario, that is, varying central densities and
 ignition geometries, has not been studied in detail, we used pure
 deflagration models of a previous parameter study dedicated to Type Iax
 supernovae as initial models to investigate the gravitationally
 confined detonation scenario. We aim to judge whether this channel can
 account for one of the many subgroups of Type Ia supernovae, or even
 normal events. To this end, we employed a comprehensive pipeline for
 three-dimensional Type Ia supernova modeling that consists of
 hydrodynamic explosion simulations, nuclear network calculations, and
 radiative transfer. The observables extracted from the radiative
 transfer are then compared to observed light curves and spectra.  The
 study produces a wide range in masses of synthesized $^{56}$Ni ranging
 from $0.257$ to $1.057\,M_\odot$, and, thus, can potentially account
 for subluminous as well as overluminous Type Ia supernovae in terms of
 brightness. However, a rough agreement with observed light curves and
 spectra can only be found for 91T-like objects.  Although several
 discrepancies remain, we conclude that the gravitationally confined
 detonation model cannot be ruled out as a mechanism to produce
 91T-like objects. However, the models do not provide a good explanation
 for either normal Type Ia supernovae or Type Iax supernovae.}

\keywords{
  supernovae: individual: SN~1991T -- SN~2012Z -- supernovae: general --
  Physical data and processes: hydrodynamics -- nucleosynthesis --
  radiative transfer -- instabilities -- turbulence stars: white dwarfs
  -- methods: numerical
}
\maketitle


\section{Introduction}
\label{sec:introduction}

Despite several decades of research, the questions of the progenitor
systems of Type Ia supernovae (SNe~Ia), and their explosion mechanism,
remain open. A common basis for modeling different conceivable explosion
scenarios is the origin of SNe~Ia from the thermonuclear disruptions of
carbon-oxygen (CO, \citealp{hoyle1960a,arnett1969a}), oxygen-neon (ONe,
\citealp{marquardt2015a,kashyap2018a}), hybrid carbon-oxygen-neon (CONe,
\citealp{denissenkov2015a,kromer2015a,bravo2016a}), or
helium-carbon-oxygen \citep{pakmor2021a} white dwarf (WD) stars in a
binary system. Whether the pre-explosion WD has reached the
Chandrasekhar mass ($M_\mathrm{Ch}$) or whether it remains significantly below this
limit is not clear yet. In fact, it may well be possible that more than
one progenitor and explosion mechanism is needed to account for the
observed sample of SNe~Ia given the variety of peculiar subclasses that
have been identified \citep{taubenberger2017a,jha2017a}. The nature of
the binary companion of the exploding WD is yet another open question
since it is not clear whether it is
a nondegenerate star, for example, a main sequence, red giant, or
asymptotic giant branch star \citep{whelan1973a}, or another WD
\citep{iben1984a}. Finally, the reason for the explosion and the
combustion mechanism are unknown. The WD might explode due to an
interaction with its binary companion via accretion or due to a merger
with the latter, and the thermonuclear flame can propagate as a subsonic
deflagration or a supersonic detonation front
\cite[e.g.,][]{roepke2017a}.  These possibilities leave room for a
plethora of explosion scenarios (see \citealp{hillebrandt2013a}, for a
review) trying to explain the typical features of the bulk of so-called
normal SNe~Ia as well as the various subclasses. Here, we focus solely
on explosions of \mch WDs.

Specifically, we extend the models of asymmetric deflagrations in \mch
WDs of \citet{lach2021a}, hereafter \citetalias{lach2021a}, and trigger
detonations in late stages of the explosion. The models presented in
that study were ignited asymmetrically in a single spot, and, hence,
lead to weak deflagrations ejecting only little mass. The deflagration
ashes rush across the still bound object and collide at the opposite
side.  Therefore, they can serve as a starting point for the
investigation of the gravitationally confined detonation (GCD) scenario
first investigated by \citet{plewa2004a}. In this scenario, a detonation
is initated in the collision region leading to a healthy explosion
disrupting the whole star. The goal of our study is to investigate two main
properties of these models. First, we explore whether this
mechanism is capable of
reproducing the brightness variation of normal SNe~Ia or even extending
to subluminous events. Second, the effects of the predicted
composition inversion in comparison to 1D SN~Ia (see e.g.,
\citealp{nomoto1984a,khokhlov1993d}) models of the canonical delayed
detonation (DD) scenario \citep{gamezo2005a,roepke2007b,sim2013a} on the
predicted observables are examined. By construction, the deflagration products
remain in the central parts of the ejecta in 1D models. In 3D DD
simulations, the composition is not strictly stratified. Deflagration
ashes extend out to high ejecta velocities while material composed of
lighter elements sinks down toward the center because of the buoyancy
instabilitiy during the initial deflagration phase. In the GCD model
this effect is most strongly pronounced. A complete inversion of the
ejecta composition compared to the 1D models is observed: the
deflagration products enshroud the detonation ashes.  While
\citet{kasen2007b} conclude that the spectrum of SN~1994D can be matched
by their GCD model, \citet{baron2008a} and \citet{seitenzahl2016a} find
it difficult to bring the large abundances of iron group elements (IGEs)
in the outer layers in agreement with observations (see discussion in
Sect.~\ref{sec:explosionmodel}).

Because of the restricted spatial resolution of the regions where the
detonation potentially triggers, the mechanism for the initiation of the
detonation wave cannot be followed consistently within this study.
According to \citet{seitenzahl2009b} a resolution of at least a few 100
meters per cell is required. We emphasize that the numerical methods
used in this work are not appropriate to conclusively verify whether a
detonation is ignited or not (see Sect.~\ref{sec:detonation}).
Therefore, we only give plausibility arguments and follow a
what-if-approach to explore the consequences of an assumed detonation
initiation. Because of this uncertainty, the occurrence of a detonation
is not necessarily a contradiction to the notion that Type Iax
supernovae (SNe~Iax) may arise from weak deflagrations in \mch WDs
\citet{branch2004a,phillips2007a,kromer2013a,long2014a,fink2014a},
\citetalias{lach2021a}. Although a detonation is possible in the
collision region it is not guaranteed that it also occurs in every
realization in nature. Furthermore, we note that the GCD mechanism
relies on the absence of a spontaneous deflagration-to-detonation
transition (DDT) before the deflagration breaks out of the surface. Also
in this case, the question - whether DDTs occur in unconfined media
remains an open issue and topic of active research
(\citealp{oran2007d,woosley2009a,poludnenko2011a}, but see
\citealp{poludnenko2019a} for a recently suggested mechanism).

The paper is structured in the following way: we first discuss the GCD
explosion mechanism and its implications in detail in
Sect.~\ref{sec:explosionmodel}, and, then give a brief overview of the
employed codes and our initial setup in Sect.~\ref{sec:numerics}. This
is followed by a description of the detonation initiation mechanism and
conditions relevant for this study in Sect.~\ref{sec:detonation}.
Sect.~\ref{sec:results} then covers the results of the hydrodynamic
simulations and the postprocessing step. In Sect.~\ref{sec:rt} we
present synthetic observables and compare them to observations. Finally,
our conclusions can be found in Sect.~\ref{sec:conclusion}.

\section{Explosion model}
\label{sec:explosionmodel}

As mentioned above, in a previous study \citepalias{lach2021a}, we
investigate explosions of \mch WDs that result from ignitions of a
deflagration flame in a single spot off-center of the star as suggested
by simmering phase simulations by \citet{zingale2009a,zingale2011a} and
\citet{nonaka2012a}. Such models had been proposed to explain the class
of SNe~Iax and we confirm that a large part of the luminosity range
observed for these objects can be reproduced in the context of the
assumed explosion scenario. The overall agreement of those models with
observational data of SNe Iax (light curves and spectra) is reasonable,
but not perfect. In particular, the observables of the lowest-luminosity
members of the SN~Iax class are not satisfactorily reproduced by our
models. In the scenario of a subsonic deflagration ignited in a single
(or a few) spark(s) off-center in a \mch WD star, the energy released by
the nuclear burning does not completely unbind the object and a bound
remnant is left behind. But even multispot ignition scenarios -- also
capable of unbinding the entire WD star -- produce \nifs masses and
explosion energies reaching the faint end of normal SNe~Ia at best
\citep{fink2014a}. A DD in such models can produce brighter
events and approximately cover the range of observed normal SNe~Ia
\citep{gamezo2005a,maeda2010c,roepke2007b,seitenzahl2013a}. Observables
predicted from these models are again in reasonable agreement with data,
however, important trends and relations such as the width-luminosity
relation employed to calibrate SNe Ia as cosmological distance
indicators \citep{phillips1993a} are not fully reproduced by current
multidimensional models \citep{kasen2009a,sim2013a}. We note, however,
that it has been demonstrated that a full treatment of nonlocal
thermodynamic equilibrium (NLTE) effects is important for detailed
modeling, including at phases relevant to studying the width-luminosity
relation (e.g., \citealp{blondin2013a}, \citealp{dessart2014a},
\citealp{shen2021a}). Since multidimensional studies currently lack
full NLTE treatments further work is therefore still needed to fully
quantify the extent to which such models can agree with observations.
As mentioned above, the mechanism triggering the required spontaneous
transition of the combustion wave from deflagration to detonation
remains unclear.

For the single-spot off-center ignited \mch WD models considered by
\citetalias{lach2021a}, triggering of a detonation in a late phase of the
explosion process is also a possibility. Such scenarios have been
discussed by \citet{plewa2004a} as gravitationally confined detonations
and were later investigated in more detail by
\citet{plewa2007a,townsley2007a,roepke2007a,jordan2008a,jordan2009a,meakin2009a,seitenzahl2009c,seitenzahl2016a,garcia2016a},
and \citet{byrohl2019a}.  The idea is that the deflagration flame rises
toward the surface of the WD and sweeps around the gravitationally
bound core. Subsequently, the ashes collide on the opposite side and a
detonation is ignited in the collision region. There, an inward and
outward moving jet form and material is compressed and heated at the
head of the inward moving jet until necessary conditions for a
detonation are reached. This leads to the following problem: the less
energy released in the initial asymmetric deflagration, the less the WD
star expands and the stronger the collision of the deflagration wave at
the antipode of its ignition.  This increases the chances of triggering
a detonation. But a weak expansion of the WD prior to this moment also
implies that the detonation produces a substantial amount of
$^{56}$Ni, and,
therefore, such explosion models are very bright.  In contrast, a too
strong deflagration may pre-expand the WD so much that the antipodal
collision of the deflagration ashes is too weak to trigger a detonation.

The immediate triggering of the detonation in the GCD model only works
if the collision of ashes is strong, that is, the deflagration is weak.
This seems too restrictive to reproduce the brightness range covered by
normal SNe~Ia, especially the fainter events. Some additional dynamics
is introduced by models that take into account the pulsation phase
ensuing in the distorted, but still bound, WD after the deflagration
phase if a detonation is not triggered immediately.  The predecessor to
such models is the pulsating delayed detonation (PDD) mechanism (see
\citealp{khokhlov1991b,khokhlov1993b,hoeflich1995b}) in which a
deflagration is ignited centrally  in a one-dimensional (1D) setup that
is too weak to unbind the WD. Hence, the star subsequently pulsates and
eventually a detonation is triggered inside a mixing zone of ashes and
heated fuel.  A multidimensional variant of this is the ``pulsationally
assisted'' GCD (PGCD, \citealp{jordan2012a}). Here, the initiation of
the detonation results from the coincidence of the contraction of the
bound core during its pulsation and the collision of the ashes.  In the
pulsating reverse detonation (PRD) mechanism proposed by
\citet{bravo2006a,bravo2009a}, the detonation does not necessarily
happen near the collision spot of the deflagration waves but is caused
by an accretion shock as the burned material falls back onto the WD
core. Most probably, these detonation initiation mechanisms coexist
legitimately and highly depend on the details of the model and the
ignition conditions. Moreover, it seems reasonable to assume that there
is a continuous transition between the GCD and the PGCD mechanism (see
also \citealp{byrohl2019a}).

All model variations mentioned above, except the original 1D PDD
scenario, predict similar ejecta structures: well mixed or even clumpy
outer layers containing IGEs and intermediate mass elements (IMEs)
typical for a deflagration and a stratified inner region caused by the
detonation of the core. There is no composition inversion in the PDD,
that is, the deflagration ashes stay in the center of the ejecta, but this
is an artifact of its restriction to one spatial dimension. 

As mentioned earlier, the actual brightness of the event is determined
by the state (density) of the WD core at detonation initiation. This
depends on the pre-expansion due to the deflagration but also on the
phase of the pulsation. Models presented to date (see references above)
either result in brightnesses comparable to luminous normal SNe~Ia,
91T-like events, or transients significantly too bright to account for
most known SNe~Ia. The masses of \nifs lie in the range of ${\sim}\, 0.7$ to
$1.2\,M_\odot$. The faint end of the distribution of normal
SNe~Ia or even bright SNe~Iax, such as SN~2012Z
\citep{stritzinger2015a}, cannot be reproduced by this scenario to
date. \citet{fisher2015a} discuss the fate of the \mch scenario and
state that sparsely ignited models most probably produce overluminous
events, such as SN~1991T. The same conclusion is reached by
\citet{byrohl2019a} who study the impact of different deflagration
ignition radii with 3D hydrodynamic simulations and find that the GCD
mechanism either fails (central ignition) or produces \nifs masses in
excess of $1\,M_\odot$.

Observational properties of the GCD model were first investigated by
\citet{kasen2005a}. They focused on the Ca~II IR triplet at high
velocities observed in some SNe Ia, for example, SN~2001el, showing that this
can be explained by deflagration ashes in the outer layers of the
ejecta. In a later study, \citet{kasen2007b} present radiative transfer
(RT) calculations of a GCD model taken from \citet{plewa2007a}. Although
the model is slightly too bright, they find a rough agreement with
SN~2001el regarding the shape of the light curve and the color
evolution.  Moreover, the spectra qualitatively match those of SN~1994D,
a normal SN~Ia, and they find strong viewing angle dependencies of the
decline rate and the Si~II $\lambda$6150 line velocities.
\citet{baron2008a} have examined a PRD model and find that it matches
the typical Si~II absorption feature at maximum light in SN~1994D quite
well while the S~II ``W'' feature, a defining characteristic of normal
SNe~Ia, is totally absent. Furthermore, the large amount of IGEs in the
outer layers leads to very red colors not compatible with normal SNe~Ia.
In a later work, \citet{bravo2009b} find that due to the asymmetry in
the deflagration ashes the spectra are not too red for all viewing
angles in the PRD models from \citet{bravo2009a}. In addition, some
bright models of their sequence agree favorably with the
[$M(^{56}\mathrm{Ni})-\Delta m_{15}$]-relation of normal SNe~Ia. The GCD
model of \citet{seitenzahl2016a} yields \mni close to that observed in
SN~1991T.  They find, however, that the amount of stable IGEs in the
deflagration ashes is too large and that features of IMEs (Ca II, S II,
Si II) are too pronounced at premaximum and too weak, in contrast, at
later epochs to match SN 1991T. Finally, significant viewing angle
effects are found, especially in the bluer bands reducing the
UV-blocking effect of IGEs in the outer layers at the detonation
initiation side. These studies show that a rough agreement of the GCD
observational properties with SNe Ia can be found although there still
are several shortcomings. Therefore, a large parameter study of the GCD
scenario is necessary to judge whether some the discrepancies between
theory and observations can be remedied.


\section{Numerical methods and initial setup}
\label{sec:numerics}

The numerical methods and the initial setup are the same as in
\citetalias{lach2021a}.  Therefore, we only give a brief review here,
and refer the reader to \citetalias{lach2021a} and references therein
for more details. 

The \textsc{leafs} code employed for the hydrodynamic simulations solves
the reactive Euler equations using the piecewise parabolic finite
difference scheme of \citet{colella1984a}. It utilizes the level-set
technique \citep{sethian1999a} to capture flame fronts and two nested,
expanding grids to follow the ejecta to homologous expansion
\citep{roepke2005c}. Gravity is treated using a fast Fourier transform
based gravity solver and the Helmholtz equation of state by
\citet{timmes2000a} is employed. Nucleosynthesis yields are calculated
via the tracer particle method \citep{travaglio2004a} with the nuclear
network code \textsc{yann} \citep{pakmor2012b}.  To obtain synthetic
spectra and light curves we carry out RT simulations using the
three-dimensional (3D) Monte Carlo RT code \textsc{artis}
\citep{sim2007b,kromer2009a} allowing comparisons to be made directly
with observational data.

We used a selection of \mch deflagration models from
\citetalias{lach2021a} as initial models for GCD explosions. CO white
dwarfs with equal amounts of C and O but different central densities
$\rho_c$ between $1\times 10^9\,\si{g.cm^{-3}}$ and $6\times
10^9\,\si{g.cm^{-3}}$ were ignited in a single bubble consisting of eight
overlapping spheres with a radius of $5\,\si{km}$ (see
\citetalias{lach2021a} for a visualization) with varying offset radii
$r_\mathrm{off}$. The latter was varied between $10$ and $150\,\si{km}$
from the center of the star.  Furthermore, the central temperature is
$6\times 10^8\,\si{K}$ decreasing adiabatically to $1 \times
10^8\,\si{K}$. In contrast to our previous \citetalias{lach2021a}
deflagration study, the metallicity was kept constant at the solar value
since this parameter has shown to be of minor importance for the
synthetic observables and the deflagration strength.  Therefore, we
abbreviate the model names and label them rX\_dY, where X encodes the
offset radius (r) in km and Y the central density (d) in
$10^9\,\si{g.cm^{-3}}$. Moreover, the cell size of the inner grid
tracking the flame is ${\sim}\,2\,\si{km}$ at the beginning of the
simulation and increases to ${\sim}\,200\,\si{km}$ at detonation
initiation.  Our explosion models are summarized in
Table~\ref{tab:ejectasum}. We omit some of the \citetalias{lach2021a}
explosion models ignited at large radii since they do not extend the
range in nuclear energy release of the deflagration
$E_\mathrm{nuc}^\mathrm{def}$, and, therefore, do not introduce further
variations to our set of GCD models.


\section{Detonation initiation}
\label{sec:detonation}

The spontaneous initiation of a detonation in a WD environment is a
complicated process and the mechanism that modern supernova research is
built on is the gradient mechanism proposed by \citet{zeldovich1970a}.
The relevant properties of a hotspot are its temperature, composition,
density and size (alternatively its mass). Studies on the initiation
process have been carried out by
\citet{khokhlov1997a,niemeyer1997b,dursi2006a}, and \citet{roepke2007a},
for instance.  \citet{seitenzahl2009b} also study different functional
forms of the temperature profile in two dimensions and find that the
ambient temperature also has an impact on the initiation process. All
these studies suggest that a detonation is robustly ignited as soon as a
density of $\rho_\mathrm{crit} > 1\times 10^7\,\si{g.cm^{-3}}$ and a
temperature of $T_\mathrm{crit} > 2 \times 10^9\,\si{K}$ are reached
simultaneously on length scales of approximately $10\,\si{km}$.
Moreover, this critical length scale increases for decreasing C mass
fraction suggesting a value of at least $X(\mathrm{C}) \sim 0.4$ is
needed.

In simulations of SNe Ia the resolution is not sufficient to capture the
details of the detonation initiation process, and, thus, the studies of
GCDs mentioned in Sect.~\ref{sec:explosionmodel} ignite a detonation as
soon as critical values are reached. \citet{seitenzahl2016a}, for
example, use a rather optimistic condition of $\rho_\mathrm{crit} >
1\times 10^6\,\si{g.cm^{-3}}$ and $T_\mathrm{crit} > 1 \times
10^9\,\si{K}$ while \citet{jordan2012a} and \citet{byrohl2019a} resort
to the conservative values of $\rho_\mathrm{crit} > 1\times
10^7\,\si{g.cm^{-3}}$ and $T_\mathrm{crit} > 2 \times 10^9\,\si{K}$. In
this work, we also stuck to these conservative conditions, and, in
addition, imposed a limit for the mass fraction of fuel, that is, 
$X(\mathrm{CO}) \ge 0.8$. The detonation is then initiated in a
spherical bubble with a radius of two cell sizes by initializing a
second levelset. Since the \textsc{leafs} code tracks the ejecta via two
nested, expanding grids the resolution in the inner parts becomes worse
with ongoing expansion. The size of one cell in the detonation region is
$\sim 200\,\si{km}$ in the simulations presented here, and, therefore,
highly exceeds the critical length scale necessary for a detonation
initiation.  Furthermore, no volume burning is included in our
simulation, and, thus, its feedback on the hydrodynamic evolution is
neglected. The low resolution at late times and the lack of an actively
coupled nuclear network therefore make it impossible to definitely
predict a detonation.  However, a detailed high resolution (up to
$125\,\si{m}$ per cell) study of the collision region in the GCD
scenario has been carried out by \citet{seitenzahl2009b}. They show that
a detonation may indeed be ignited as the result of a complex interplay
of internal shocks, compression waves and Kelvin-Helmholtz instabilities
associated with an inward moving jet near the collision area.


\section{Results}
\label{sec:results}

\subsection{Hydrodynamic evolution}
\label{subsec:hydro}

\begin{figure*}[htbp]
  \centering
  \scriptsize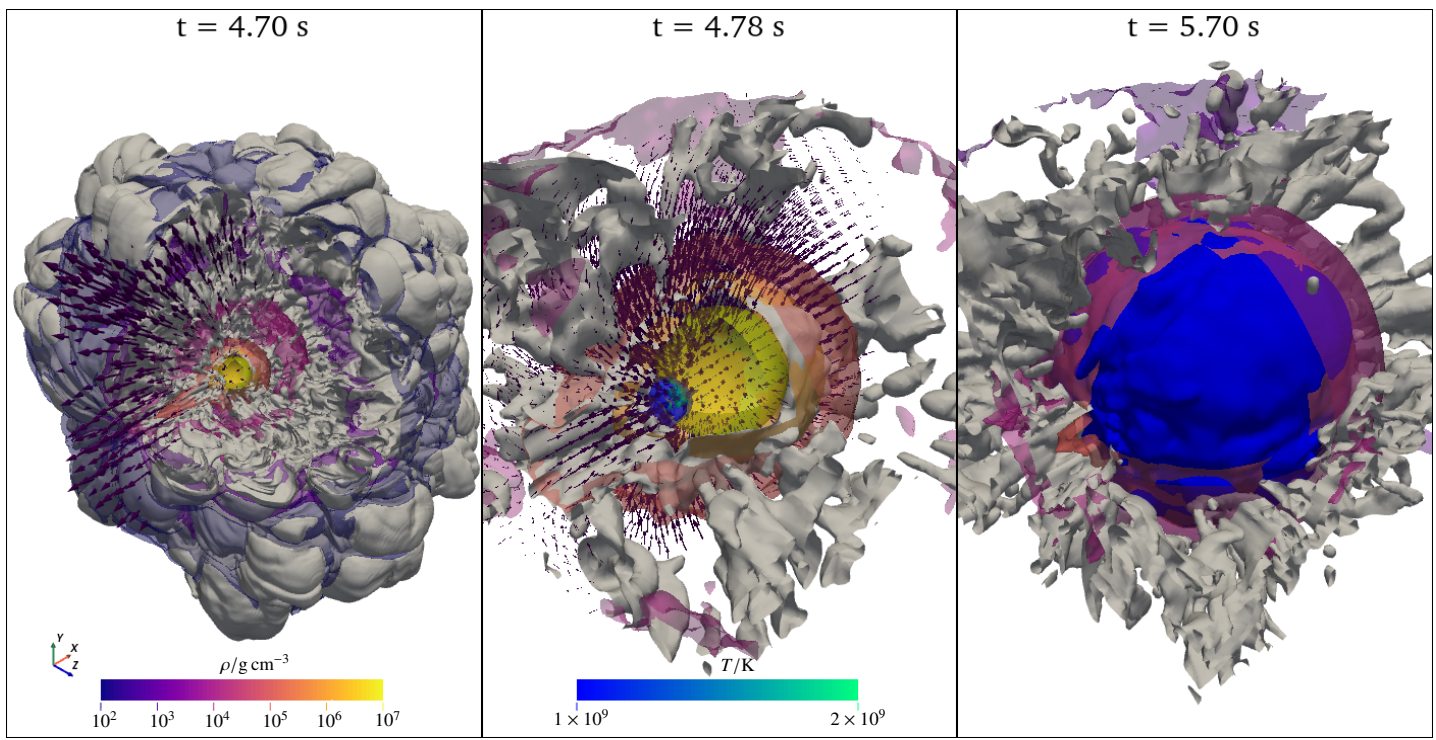
  \caption{Flame surface (gray) and density isosurfaces of 
    $\rho=10^{2},\,10^{3},\,10^{4},\,10^{5},\,10^{6},\,10^{7}\,\si{g.cm^{-3}}$
    (see colorbar) for model r60\_d2.6. The left panel shows the
    situation prior to detonation initiation at $t=4.7\,$s. The middle panel
    shows a zoom in on the core at $t=4.78$s. The hotspot with
    temperatures above $1 \times 10^{9}\,\si{K}$ is marked by the
    blue-green contour. The right panel shows the detonation front (blue
    surface) ${\sim} \, 1\,\si{s}$ after initiation. The
    illustration is not to scale.}
  \label{fig:3d}
\end{figure*}

We find that the conditions necessary for initiating a detonation in the
context of the PGCD scenario (see Sect.~\ref{sec:detonation}) are
achieved in most of the models from \citetalias{lach2021a}. Only the
models ignited at $r_\mathrm{off}=10\,\si{km}$ with central densities in
excess of $2.6 \times 10^{9}\,\si{g.cm^{-3}}$ and the rigidly
rotating models do not reach the detonation initiation conditions in the
collision area.  In high central density models (r10\_d3.0\_Z,
r10\_d4.0\_Z, r10\_d5.0\_Z, and r10\_d6.0\_Z from
\citetalias{lach2021a}), the deflagration is too powerful, that is, high
values of $E_\mathrm{nuc}^{\mathrm{def}}$, and the pre-expansion of the
star is very strong. Hence, the high densities needed for the initiation
of a detonation are not reached. For the highest central densities the
expansion is so strong that the ashes hardly collide at the antipode
(see also discussion in \citetalias{lach2021a}). The
rigidly rotating models of \citetalias{lach2021a} do not detonate although their nonrotating
counterpart, Model r60\_d2.0, detonates via the PGCD mechanism. This
supports the work of \citet{garcia2016a} stating that rotation breaks
the symmetry and leads to a weaker focusing of the colliding
deflagration products. In summary, high temperatures and densities
sufficient to initiate a detonation are found in 23 of the 29
single-spark ignition models presented in \citetalias{lach2021a}. We,
however, only simulated the detonation phase for eleven models to
capture the whole variety in explosion energies and production of
$^{56}$Ni, respectively. The models omitted are very weak deflagrations
resulting in very bright events if a detonation was taken into account.

\begin{figure}[htbp]
  \centering
  \includegraphics{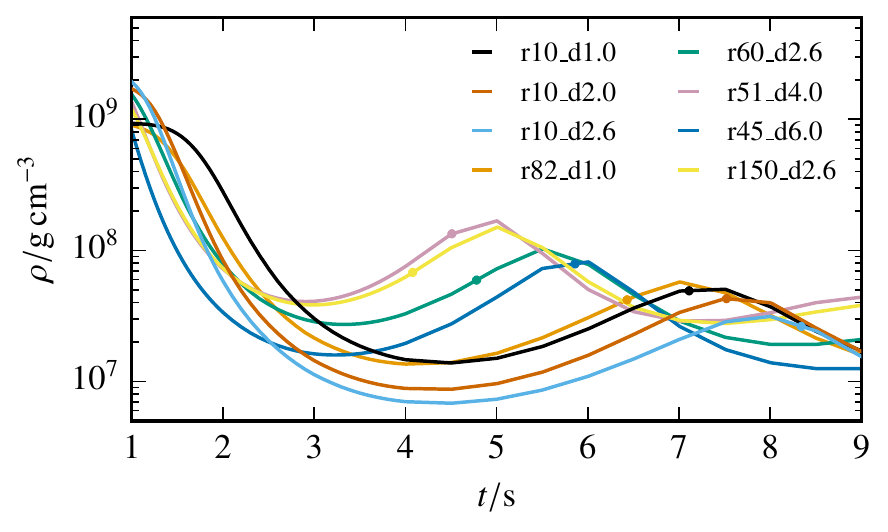}
  \caption{Central density over time for a representative selection of
  models. The scatter points indicate the time of the detonation
initiation. }
  \label{fig:pulsation}
\end{figure}

\begin{figure*}[htbp]
  \centering
  \includegraphics[width=0.97\textwidth]{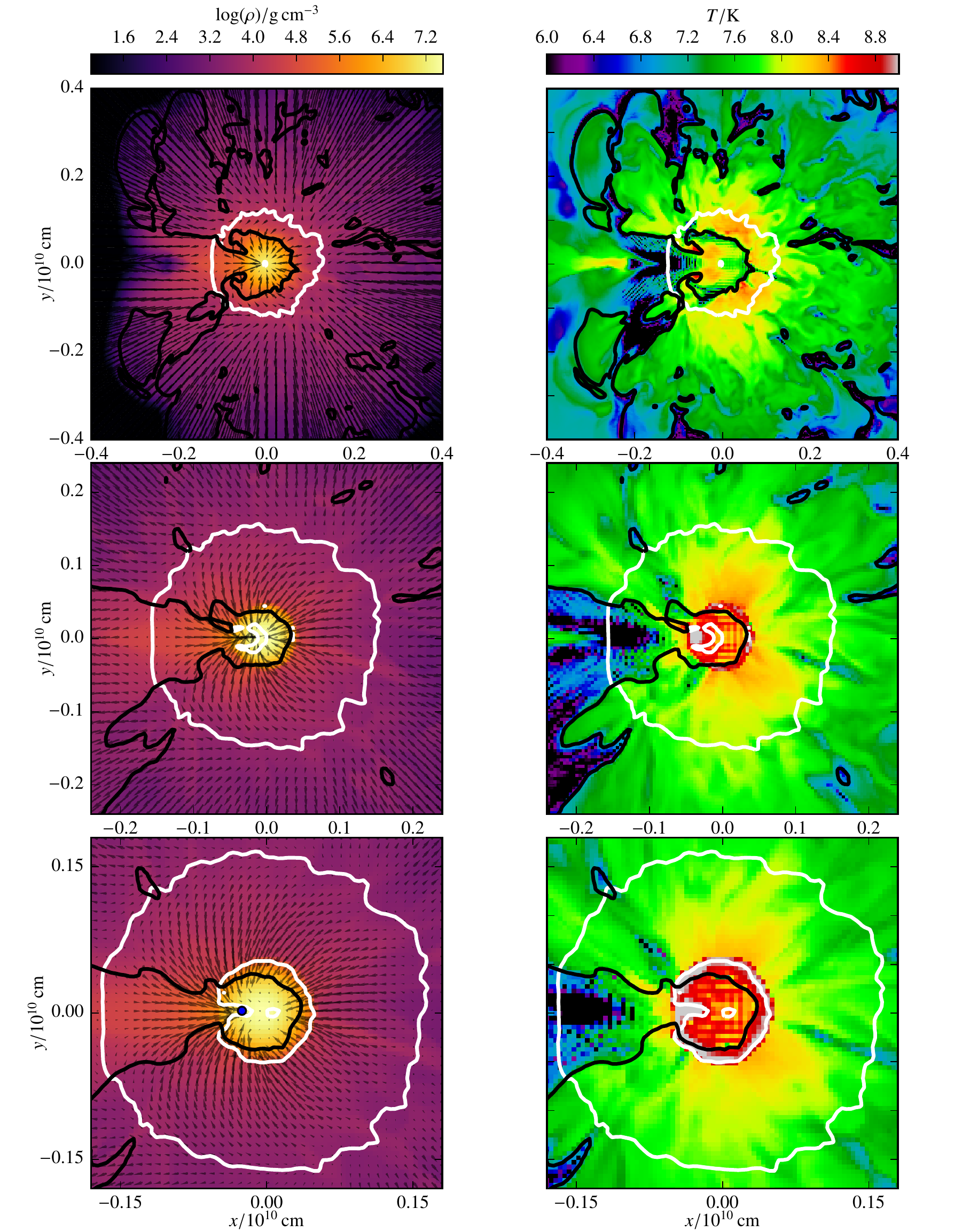}
  \caption{Slices of the $x$-$y$-plane at $z=0$ showing the density
    (left column) and the temperature (right column) of Model r10\_d2.6
    for three points in time: $6.21\,$s, $7.71\,$s and $8.31\,$s (from
    top to bottom). The white contour marks the zero level of the radial
    velocity, i.e., it separates regions of expansion and contraction and the black
    contour indicates $X(\mathrm{CO})=0.8$. The blue point in the bottom left panel shows
  the location of the detonation initiation.}
  \label{fig:gcdslice}
\end{figure*}

The evolution of the explosions is similar to that described in earlier
works
\citep{plewa2007a,townsley2007a,roepke2007a,seitenzahl2016a,byrohl2019a}:
after the ignition of the deflagration, the flame buoyantly rises to the
surface of the star accelerated further by turbulent motions
(Rayleigh-Taylor and Kelvin-Helmholtz instabilities). Subsequently, the
hot ashes race across the surface of the bound core and clash at the
side opposite to the deflagration ignition. A stagnation region, that is, a
region of almost zero velocity, then develops and two jets are launched
inward and outward. They are sustained by the ongoing inflow of material
from all lateral directions. A 3D visualization of this process is shown
in Fig.~\ref{fig:3d}. As the jet moves inward it compresses and heats
matter ahead of it until $\rho_\mathrm{crit}$ and $T_\mathrm{crit}$ are
reached. These conditions are achieved during the first pulsation phase
of the core, and, therefore, all our models belong to the PGCD scenario
\citep{jordan2012a}. Fig.~\ref{fig:pulsation} shows the central density
as a function of time where the detonation initiation of each model is
marked with a filled circle. In general, for the weak deflagrations the
detonation already occurs before the maximum compression of the core is
reached and shifts to later points in time for the stronger ones marking
the transition from the PGCD to the GCD scenario. Model r10\_d2.6 even
ignites during the onset of the expansion phase after the maximum
central density was reached in the first pulsation cycle (see
Fig.~\ref{fig:pulsation}).  This special situation is shown in
Fig.~\ref{fig:gcdslice}.  At $t=6.21\,$s (upper row) the core is
contracting (inside the white contour) and the inward moving jet pushes
material toward the center, and, thus,  compresses and heats it. The
temperature in the compression region is still low around $1 \times
10^{8}\,\si{K}$. About $1.5\,$s later, the core has reached its maximum
density and begins to expand again (see the second white contour
emerging in the dense core, middle row). The temperature is already near
$1 \times 10^{9}\,\si{K}$. Right before the initiation of the detonation
at $8.31\,\si{s}$ the infalling material further heats the expanding
innermost parts of the core. This heating is strongest where the inward
directed jet penetrates into the core until the detonation conditions
are reached. The initiation location is indicated by a scatter point in
the last row of Fig.~\ref{fig:gcdslice}. The detonation then burns the
remaining CO fuel at supersonic velocities in less than $\sim
0.5\,\si{s}$ and the whole star is disrupted. 

\begin{table*}[htbp]
  \centering
  \caption{Summary of the main properties of the ejected material and
  the initial conditions.}
 \begin{tabularx}{\textwidth}{c @{\extracolsep{\fill}} cccccccccccc}
    \toprule
    model & $\rho_c$ & $r_\mathrm{ign}$ & $E_\mathrm{nuc}^\mathrm{def}$ & $E_\mathrm{nuc}$ &
    $t_\mathrm{det}$ & $ \rho_c^\mathrm{det}$& $M(^{56}$Ni) & $M_\mathrm{IGE}$ & $M_\mathrm{IME}$ &
    $E_\mathrm{kin,ej}$  \\ 
    & \small $(10^{9}\,\si{g.cm^{-3}})$ & \small(km) & \small$(10^{50}\si{erg})$ &
    \small$(10^{51}\si{erg})$ & \small(s) & \small $(10^{7}\,\si{g.cm^{-3}})$ & \small(\msun) & \small(\msun) & \small(\msun) &
    $(10^{51}\si{erg})$ \\
    \midrule
    r10\_d1.0 & 1.0 & 10  & 1.98 & 1.70 & 7.11 & 5.12 & 0.596 & 0.662 & 0.521 & 1.24 \\     
    r10\_d2.0 & 2.0 & 10  & 2.81 & 1.69 & 7.52 & 4.37 & 0.532 & 0.612 & 0.561 & 1.21 \\
    r10\_d2.6 & 2.6 & 10  & 3.15 & 1.61 & 8.34 & 2.96 & 0.257 & 0.328 & 0.782 & 1.14 \\
    r82\_d1.0 & 1.0 & 82  & 2.31 & 1.63 & 6.43 & 4.12 & 0.592 & 0.657 & 0.507 & 1.17 \\
    r65\_d2.0 & 2.0 & 65  & 2.38 & 1.74 & 5.65 & 4.86 & 0.695 & 0.773 & 0.451 & 1.24 \\
    r60\_d2.6 & 2.6 & 60  & 1.93 & 1.79 & 4.78 & 5.86 & 0.793 & 0.876 & 0.386 & 1.30 \\
    r57\_d3.0 & 3.0 & 57  & 1.86 & 1.85 & 4.97 & 7.00 & 0.850 & 0.936 & 0.346 & 1.36 \\
    r51\_d4.0 & 4.0 & 51  & 1.29 & 1.96 & 4.51 & 13.4 & 1.057 & 1.164 & 0.175 & 1.48 \\
    r48\_d5.0 & 5.0 & 48  & 1.67 & 1.96 & 5.05 & 11.7 & 0.916 & 1.024 & 0.288 & 1.47 \\
    r45\_d6.0 & 6.0 & 45  & 2.19 & 1.93 & 5.86 & 8.80 & 0.760 & 0.876 & 0.412 & 1.45 \\
    r150\_d2.6 & 2.6 & 150 & 1.75 & 1.81 & 4.09 & 6.73 & 0.914 & 0.999 & 0.289 & 1.29 \\
    \bottomrule
  \end{tabularx}
  \label{tab:ejectasum}
\end{table*}

We emphasize that although many of our simulations robustly ignite a
detonation, this is not proof for the mechanism since the resolution in
the area of interest is not fine enough to track the turbulent motions
and shock waves inside the collision region (see
\citealp{seitenzahl2009b}) due to the expanding grid approach to track
the ejecta. Moreover, we did not employ a reaction network in parallel
with the hydrodynamics to capture the nuclear energy generation outside
the flame. Therefore, the detonation location as well as
$t_\mathrm{det}$ can be subject to slight variations within a specific
model which may introduce some more diversity among the presented PGCD
models. The energetic and nucleosynthesis yields as well as the
synthetic observables of our simulations, however, are expected to
largely capture the characteristics of these explosions. 

\subsection{Energetics and nucleosynthesis yields}
\label{subsec:energy}

The total nuclear energy $E_\mathrm{nuc}$ released during the explosion
is sufficient to unbind the whole WD in all simulations of this work.
The disruption of the whole star was verified by checking that the
kinetic energy of each cell exceeds its gravitational energy at the end
of the simulations ($t=100\,\si{s}$). The ejecta expand homologously.
While $E_\mathrm{nuc}$ ranges from 1.61 to $1.96 \times
10^{51}\,\si{erg}$ the energy released in the deflagration
$E_\mathrm{nuc}^\mathrm{def}$ lies between 1.29 and $3.15\,\times
10^{50}\,\si{erg}$ (see Table~\ref{tab:ejectasum}) which corresponds to
24.1 to 61.4\% of the initial binding energy $E_\mathrm{bind}$ of the
star. There is a clear trend that for an increasing deflagration
strength ($E_\mathrm{nuc}^\mathrm{def}$) the time until detonation
$t_\mathrm{det}$ increases and the central density $\rho_\mathrm{det}$
at $t_\mathrm{det}$, $E_\mathrm{nuc}$ and with it the total kinetic
energy $E_\mathrm{kin}$ decrease.  This behavior is well in line with
previous studies of the GCD scenario (see Sect.~\ref{sec:introduction}).
A little scatter is added to these trends by the different initial
central densities of the models. As an extreme example, Model r45\_d6.0
and Model r82\_d1.0 release a comparable amount of nuclear energy in the
deflagration but differ significantly in their total energy release
$E_\mathrm{nuc}$, and, especially in the amount of \nifs produced. The
high initial central density of Model r45\_d6.0 also leads to a high
value for $\rho_\mathrm{det}$ of $8.8 \times 10^7\,\si{g.cm}^{-3}$ more
than twice the value found for Model r82\_d1.0. Moreover, the total
amount of matter at high densities above $10^7\,\si{g.cm}^{-3}$ at the
detonation initiation sums up to $0.98\,M_\odot$ in Model r45\_d6.0 and
to $0.64\,M_\odot$ in Model r82\_d1.0, respectively. 

The parameter driving the luminosity of SNe Ia is the mass of \nifs
ejected since it deposits energy into the expanding material via its
decay to $^{56}$Co, and, subsequently, to $^{56}$Fe. The model sequence
presented here covers a wide range of luminosities resulting from \mni
ranging between $0.257$ and $1.057\,M_\odot$. This includes $^{56}$Ni
masses appropriate for faint normal SNe Ia (see e.g.
\citealp{stritzinger2006b,scalzo2014a}) but also the bright end of the
SN Iax subclass (e.g., SN~2012Z ejecting $\sim 0.2-0.3\,M_\odot$ of
\nifs and $\sim M_\mathrm{Ch}$ in total \citealp{stritzinger2015a} or
SN~2011ay \citealp{szalai2015a}) and 02es-like objects (see SN~2006bt
also ejecting $\sim 0.2-0.3\,M_\odot$ of \nifs \citealp{foley2010c}). On
the bright end, luminous normal SNe Ia, the slightly overluminous
91T-like objects \citep{filippenko1992a} and the transitional events
like SN~2000cx \citep{li2001b} can also be accounted for by our models
in terms of \mni ejected. 

The result that models ignited at $r_\mathrm{off}=10\,\si{km}$ do
detonate pushes the transition between PGCDs and failed detonations,
that is, pure deflagrations, to lower ignition radii compared to the works
of \citet{fisher2015a} and \citet{byrohl2019a}. These studies conclude
that below ${\sim}\,16\,\si{km}$ a detonation initiation becomes
unlikely, and, thus, the GCD mechanism produces mostly bright
explosions. Indeed, the faintest model in our study, Model r10\_d2.6,
exhausts the viability of the PGCD mechanism since critical conditions
for a detonation initiation are only reached at the very end of the
contraction phase. However, subluminous SNe~Ia resulting from the GCD
scenario should still be considered rare events as pointed out by
\citet{byrohl2019a}. They extract a probability density function (PDF)
for the ignition radius using data from \citet{nonaka2012a} to interpret
their results. The PDF peaks at approximately $50\,\si{km}$ and suggests
that ignition below ${\sim}\,20\,\si{km}$ and above
${\sim}\,100\,\si{km}$ is a scarce event.  In detail, they state that
only $2.2\,$\% of \mch explosions are ignited below
$r_\mathrm{off}=16\,\si{km}$.

\citet{flors2020a} examine features of stable Ni, mostly in the form of
$^{58}$Ni, and Fe in late-time spectra of SNe Ia. They then compare
their measured Ni to Fe ratios ($M_\mathrm{Ni}/M_\mathrm{Fe}$) to
theoretical models and conclude that most SNe Ia probably originate from
sub-\mch WDs. In detail, they find that
$M_{^{58}\mathrm{Ni}}/M_{^{56}\mathrm{Ni}}<6\%$ for sub-\mch models (see
references in \citealp{flors2020a}) and larger for \mch explosions. The
reason for this is that more neutron-rich isotopes are favored in
nuclear statistical equilibrium at high densities which are reached in
\mch WDs only. We find values of
$M_{^{58}\mathrm{Ni}}/M_{^{56}\mathrm{Ni}}$ between 2.4\% and 5.1\%
making it hard to distinguish the PGCD models\footnote{The full set of
  isotopic yields and also the angle averaged optical spectral time
  series and UVOIR bolometric light curves calculated in the radiative
  transfer simulations will be published on \textsc{HESMA}
  \citet{kromer2017a}.} from sub-\mch explosions based on this ratio.
  Moreover, \citet{flors2020a} state that
  $M_{^{57}\mathrm{Ni}}/M_{^{56}\mathrm{Ni}}$ is expected to lie above
  2\% and $M_{^{54,56}\mathrm{Fe}}/M_{^{56}\mathrm{Ni}}$ above 10\% for
  \mch models. Again, the nucleosynthesis yields of our simulations
  predict lower values of
  $M_{^{57}\mathrm{Ni}}/M_{^{56}\mathrm{Ni}}=1.7-2.4\%$ and
  $M_{^{54,56}\mathrm{Fe}}/M_{^{56}\mathrm{Ni}}=1.8 - 14.6\%$. The
  obvious reason for this finding is that PGCD models exhibit
  characteristics of a deflagration at high densities and a detonation
  of a sub-\mch CO core. Finally, the significant amounts of IMEs
  produced in Model r10\_d2.6 even exceed the mass of IGEs.

\begin{figure}[htbp]
  \centering
  \includegraphics{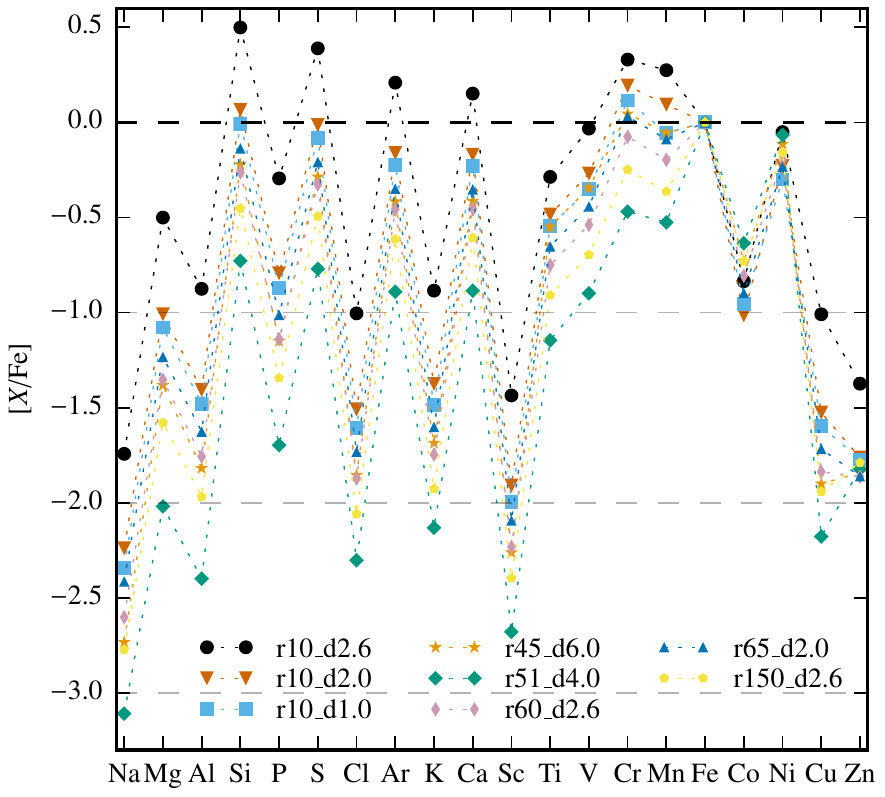}
  \caption{Elemental abundances over Fe in relation to their solar value
  for a few representative models of the parameter study.}
  \label{fig:solratio}
\end{figure}

The elemental ratios to Fe compared to their solar value [X/Fe]
(isotopes decayed to $2\,$Gyr) also include characteristics of both \mch
deflagrations and detonations in low-mass WDs. First, supersolar values
of [Mn/Fe] are achieved in models with a strong deflagration, for
instance,
Model r10\_d2.6 and r10\_d2.0 (see Fig.~\ref{fig:solratio}). This is in
agreement with the commonly accepted fact that \mch explosions are
responsible for a substantial amount of Mn in the Universe
\citep{seitenzahl2013b,lach2020a}. For decreasing deflagration strength
the contribution of the detonation becomes larger and [Mn/Fe] falls
below the solar value. Second, we observe high values of [Cr/Fe] in some
explosions which is a direct imprint of low-mass ($\lesssim
0.8\,M_\odot$) CO detonation models (compare also Fig.~3 in
\citealp{lach2020a}) but special for \mch models. Third, the usual
overproduction of stable Ni in \mch deflagrations is suppressed by the
detonation yields. Finally, Fig.~\ref{fig:solratio} also shows signs of
the strong odd-even effect in the detonation IME yields smoothed out by
the deflagration products (compare again Fig.~3 in \citealp{lach2020a}).
The $\alpha$-elements Si, S, Ar and Ca are even synthesized in
supersolar amounts in the core detonation of Model r10\_d2.6.  The rest
of the models show rather low values of [$\alpha$/Fe] more typical for
\mch deflagrations. In summary, the nucleosynthesis yields of faint
(r10\_d2.6) to intermediately bright (r10\_d2.0) models show
characteristics of low-mass CO detonations as well as \mch explosions.
As the deflagration strength decreases, the total elemental yields more
and more resemble those of a pure CO detonation in a WD of approximately
$1\,M_\odot$. Nevertheless, the pollution of the outer layers with
deflagration products still has some impact on the observables (see
Sect.~\ref{sec:rt}).

\subsection{Ejecta structure}
\label{subsec:ejecta}

\begin{figure*}[htbp]
  \centering
  \includegraphics[width=0.95\textwidth]{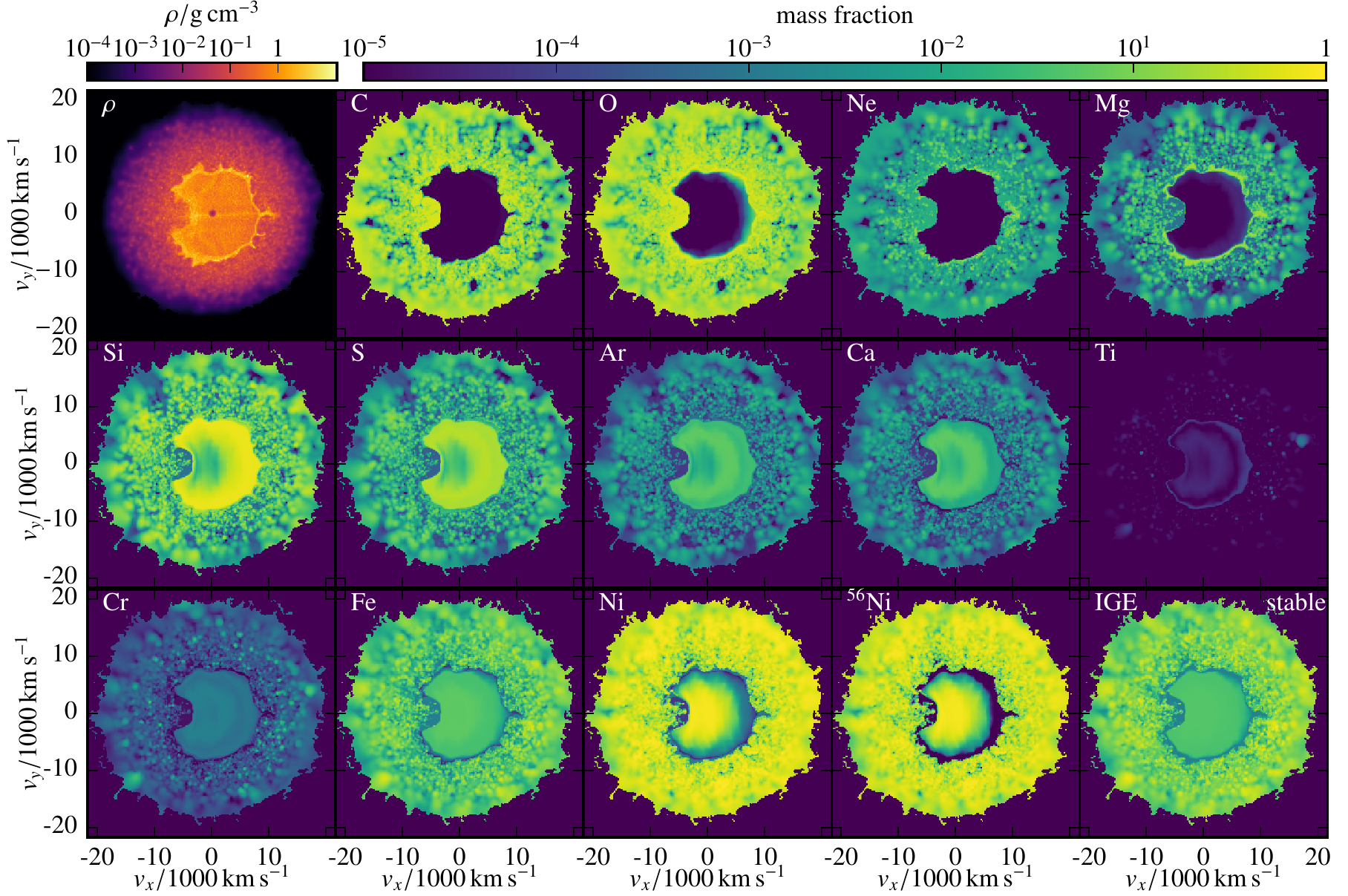}
  \caption{Color-coded density and elemental abundance slices in the $v_x - v_y$-plane for
  Model r10\_d2.6.}
  \label{fig:abundslicer10d26}
  \vspace{0.3cm}
  \centering
  \includegraphics[width=0.95\textwidth]{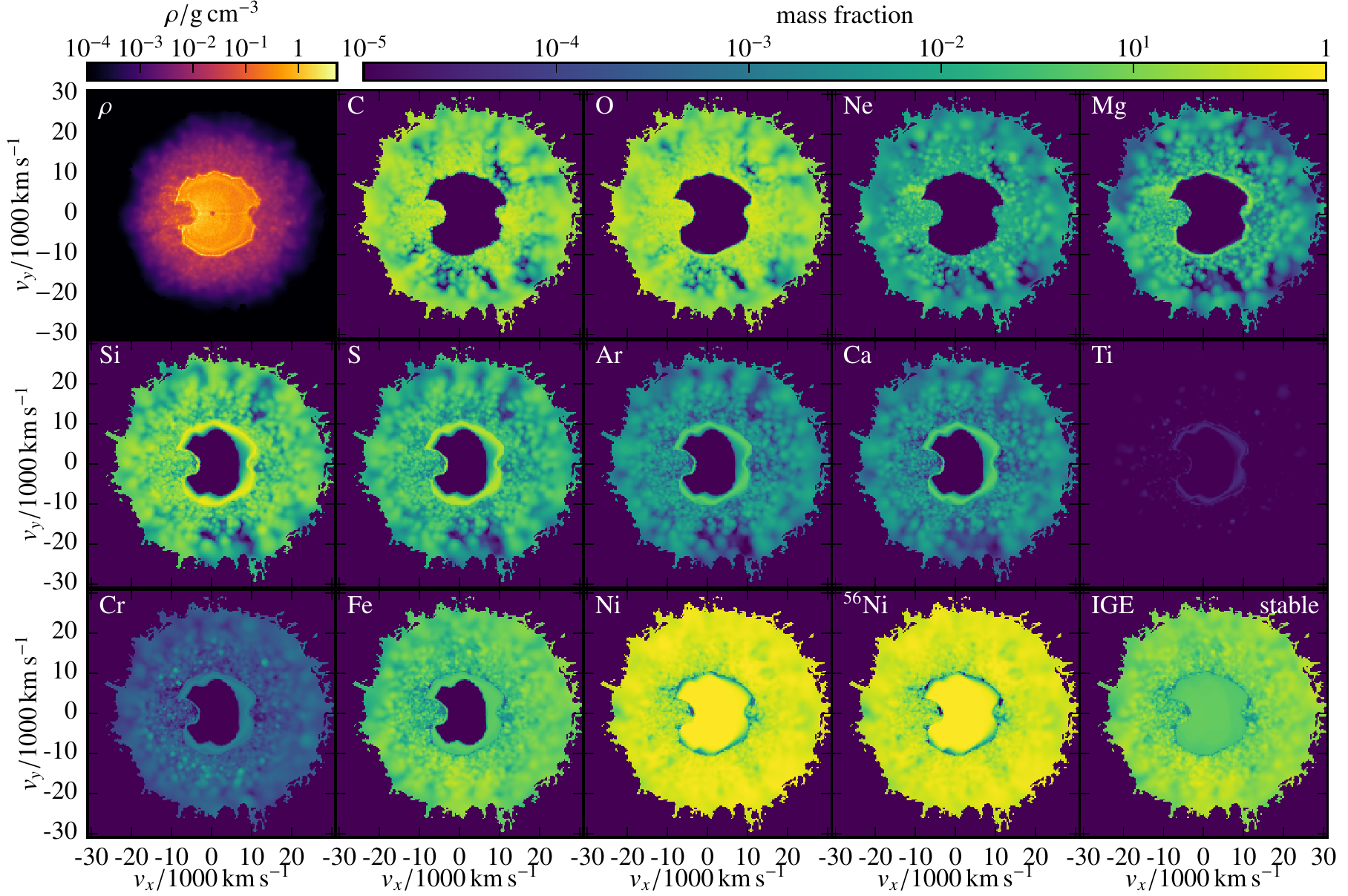}
  \caption{Same as Fig.~\ref{fig:abundslicer10d26}, but for Model r60\_d2.6.}
  \label{fig:abundslicer60d26}
\end{figure*}

\begin{figure*}[htbp]
  \centering
  \includegraphics[width=0.95\textwidth]{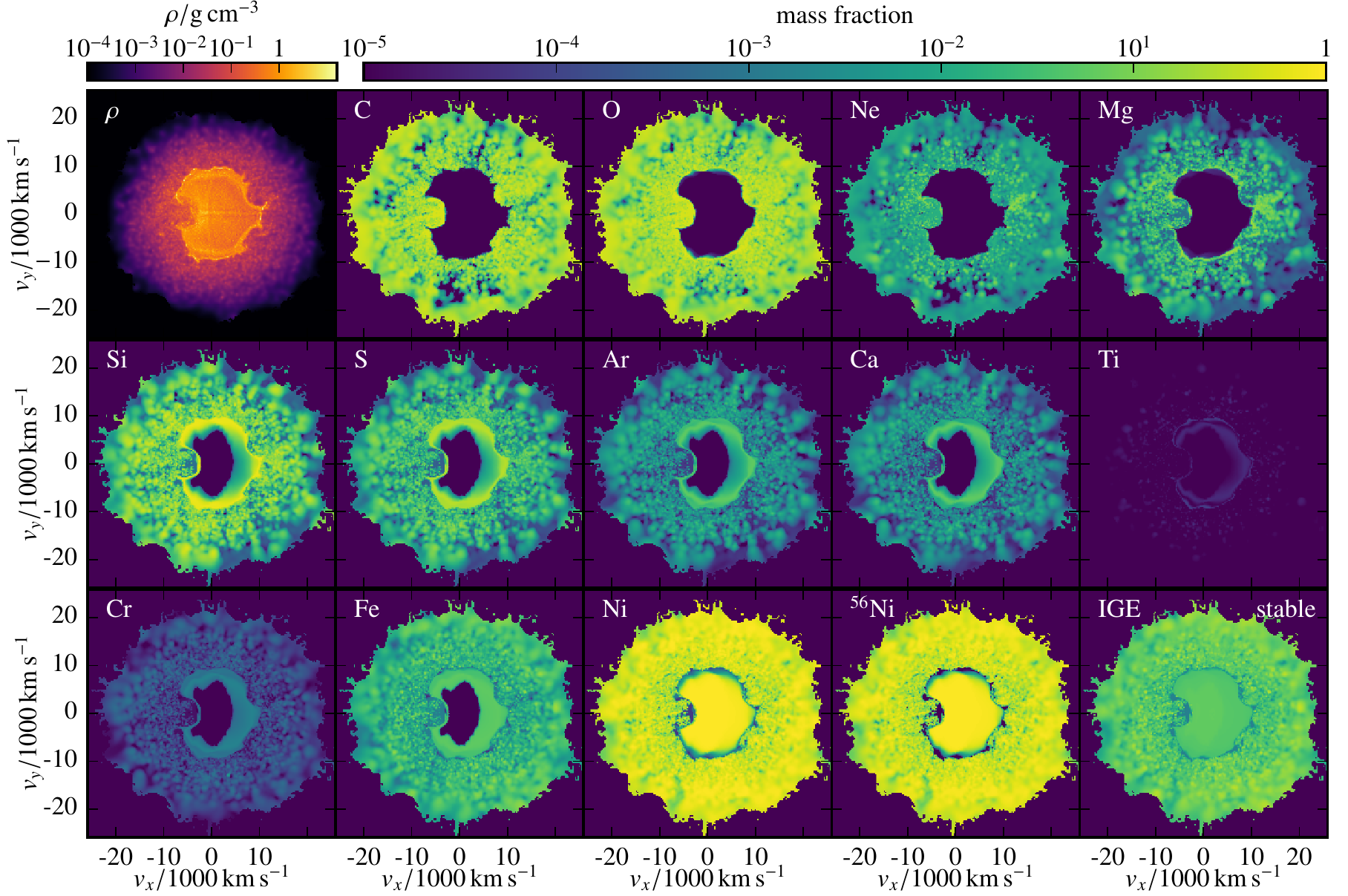}
  \caption{Same as Fig.~\ref{fig:abundslicer10d26}, but for Model r82\_d1.0.}
  \label{fig:abundslicer82d1}
  \vspace{0.3cm}
  \centering
  \includegraphics[width=0.95\textwidth]{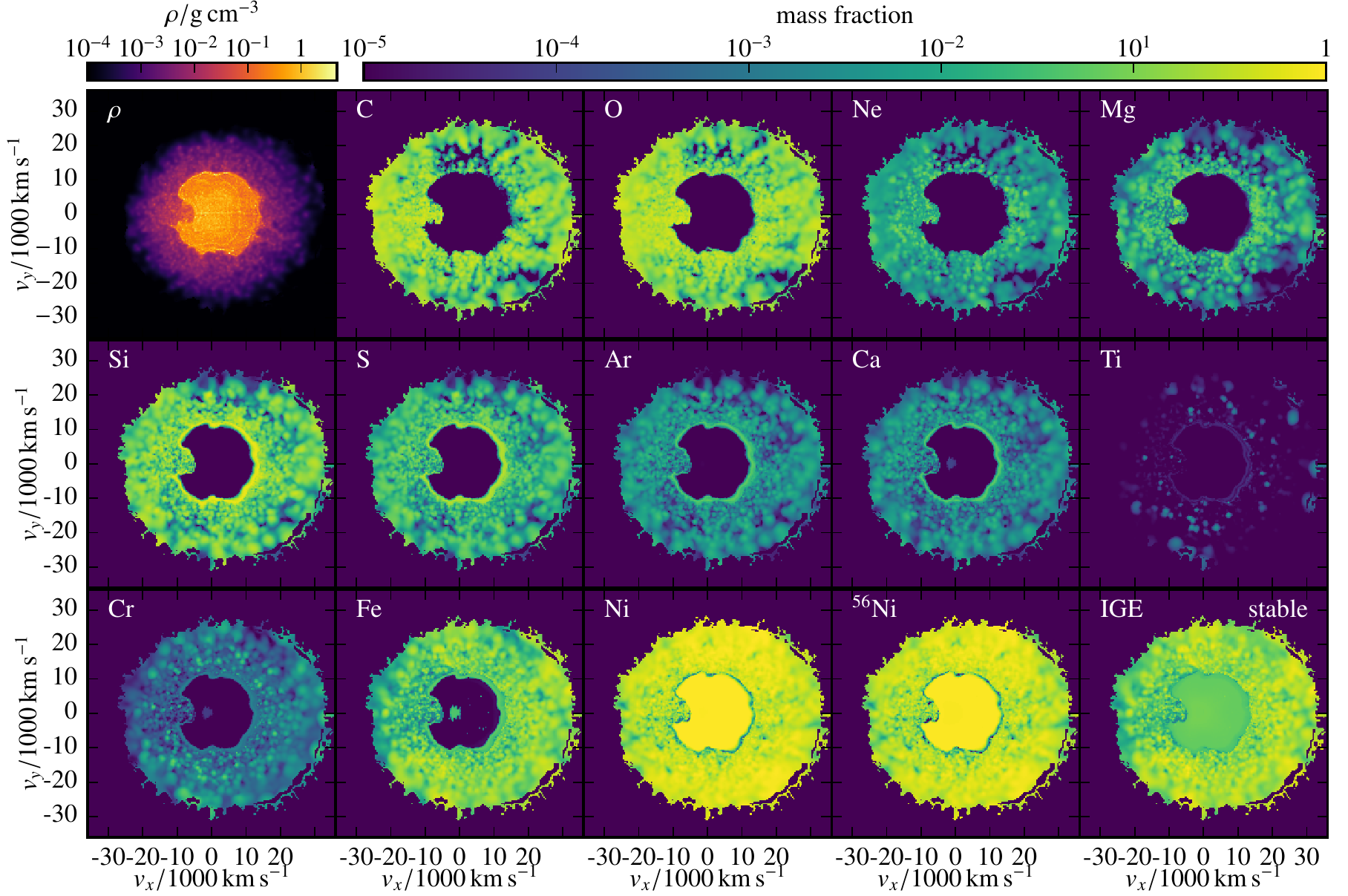}
  \caption{Same as Fig.~\ref{fig:abundslicer10d26}, but for Model r51\_d4.0.}
  \label{fig:abundslicer51d4}
\end{figure*}

The expelled material is spherically symmetric on large scales,
especially the deflagration ashes which reach maximum velocities from
${\sim}\, 19\,000\,\si{km.s}^{-1}$ (r10\_d2.6, see
Fig.~\ref{fig:abundslicer10d26}) up to ${\sim}\, 29\,000\,\si{km.s}^{-1}$
(r51\_d4.0, see Fig.~\ref{fig:abundslicer51d4}).  The shell of well
mixed deflagration products extends down to ${\sim}\,
8\,000\,\si{km.s}^{-1}$ in Model r10\_d2.6 and ${\sim}\,
11\,000\,\si{km.s}^{-1}$ for Model r51\_d4.0. The outer boundary of the
detonation products, primarily consisting of IMEs, is not exactly
spherical but slightly elliptical and more elongated along the $y$-axis
(perpendicular to the line connecting the deflagration and the
detonation initiation). This structure is more pronounced in Model
r82\_d1.0 (see Fig.~\ref{fig:abundslicer82d1}) and only very subtle in
the most energetic explosion, that is, the strongest detonation, r51\_d4.0.
Moreover, the detonation region exhibits a cusp of deflagration products
on the detonation side (left-hand side) disturbing the spherical
structure.  This is the region of the jet of ashes penetrating into the
core until the detonation conditions are reached. 

A comparison of the ejecta structure to Fig.~4 of
\citet{seitenzahl2016a} reveals an important difference: In their work
the detonating core is not centered inside the deflagration ashes. This
means that the envelope of burned material is thicker on the
deflagration ignition side than on the detonation side leading to
significant viewing angle effects. In our work the detonation products
lie relatively well centered inside the shell of deflagration ashes (see
Figs.~\ref{fig:abundslicer10d26}, \ref{fig:abundslicer60d26},
\ref{fig:abundslicer82d1}, \ref{fig:abundslicer51d4}). Only a very close
inspection of Model r51\_d4.0 (brightest) and Model r10\_d2.6 (faintest)
reveals that the deflagration envelope is slightly thicker on the
deflagration side in Model r51\_d4.0 than on the antipode. The opposite
holds for Model r10\_d2.6the deflagration envelope is slightly thicker
on the deflagration side in Model r51\_d4.0 than on the antipode. The
opposite holds for Model r10\_d2.6.  In addition to this minor
asymmetry, viewing angle effects can be expected to originate from the
irregularly shaped core detonation. This difference between our models
and the model of \citet{seitenzahl2016a} is very likely due to the
distinct detonation initiation conditions. Since \citet{seitenzahl2016a}
use optimistic values of $\rho=10^6\,\si{g.cm^{-3}}$ and
$T=10^9\,\si{K}$ the detonation is ignited very early
($t_\mathrm{det}=2.37\,\si{s}$) almost immediately after the ashes have
collided. Therefore, the amount of ashes on the detonation side has not
piled up significantly and the detonation also ignites more off-center
than found in this study. In our models the detonation only ignites
after a prolonged period of contraction and the penetration of a jet
into the core ($t_\mathrm{det} > 4.5\,\si{s}$) providing enough time for
the deflagration products to engulf the core homogeneously.

\begin{figure}[htbp]
  \centering
  \includegraphics{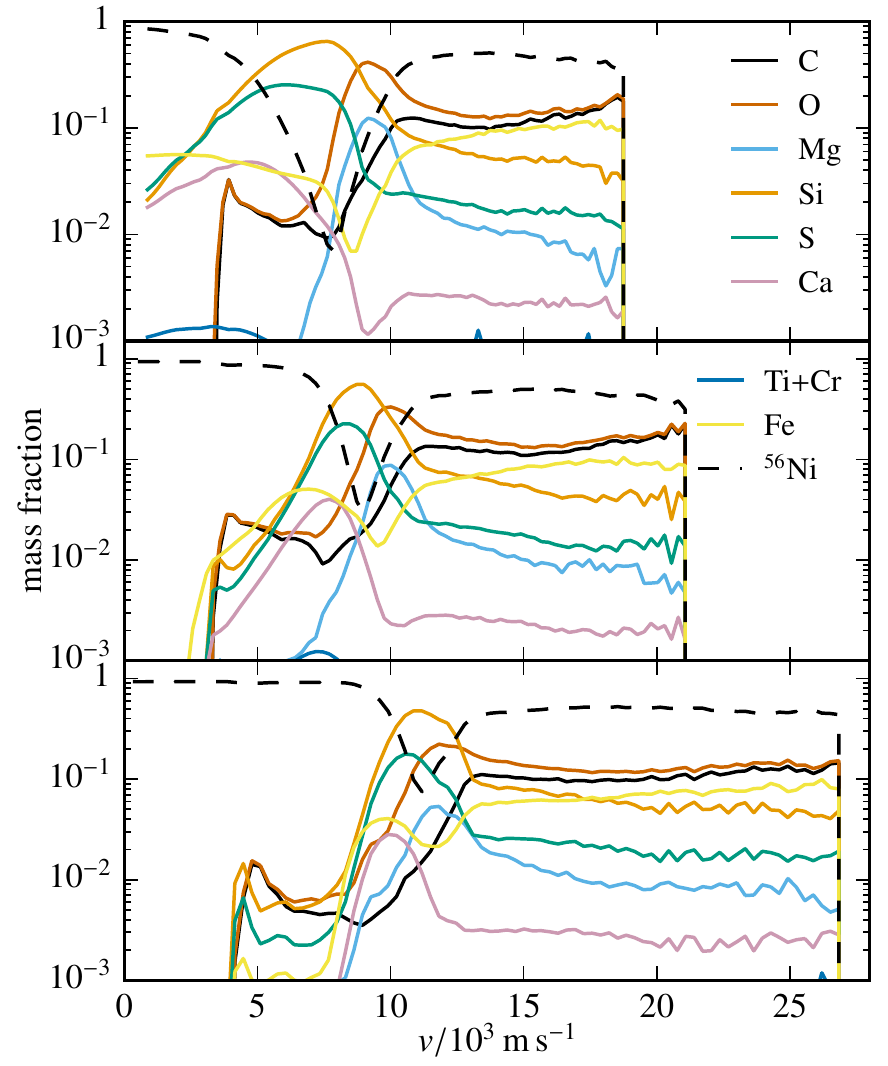}
  \caption{1D-averaged velocity profiles for C, O, Mg, Si, S, Ca, Ti+Cr,
  Fe and \nifs. The upper panel shows data for Model r10\_d2.6, the mid panel
Model r10\_d2.0 and the lower panel Model r57\_d3.0.}
  \label{fig:velocityprofile}
\end{figure}

The 1D-averaged velocity profiles of various elements and \nifs are
shown in Fig.~\ref{fig:velocityprofile} for a faint (r10\_d2.6), a
moderately bright (r10\_d2.0) and an overluminous (r51\_d3.0) model. All
explosions are characterized by mixed deflagration products at high
velocities dominated by \nifs. The second most abundant elements are
unburned C and O followed by Fe and Si. Less abundant IMEs present in
the outer ejecta are S, Mg, and Ca while the mass fractions of  Ar, Ti
and Cr stay below $10^{-3}$. Going to lower velocities, the IMEs
individually (from light to heavy) rise to a maximum and Fe and \nifs
exhibit a local minimum. This stratified region was burned in the
detonation at rather low densities via incomplete silicon burning. The
innermost regions are then dominated by \nifs. The central density at
detonation in Model r10\_d2.6, however, is so low that even a
significant fraction of IMEs is present at low velocities. 

From the ejecta structure it is evident that all of these models will
display IGEs in their early-time spectra. This characteristic is in
agreement with SNe Iax \citet{foley2013b,jha2017a} as well as the
overluminous 91T-like events. Normal SNe Ia, however, are dominated by
IMEs around maximum light. Moreover, the fact that stable IGE are
primarily found in the outer layers at high velocity neither matches the
properties of 91T-like \citep{sasdelli2014a,seitenzahl2016a} objects nor
normal SNe~Ia \citep{mazzali2007a}. The distribution of stable IGEs is
also a problem for the method used in \citet{flors2020a} (see also
Sect.~\ref{subsec:energy}) to distinguish between \mch and sub-\mch mass
models since only a small fraction of stable IGEs will be visible in
late-time spectra. 

\section{Synthetic observables}
\label{sec:rt}

To obtain synthetic spectra and light curves for our sequence of PGCD
models we carried out time-dependent 3D Monte-Carlo RT simulations using
the \textsc{artis} code \citep{sim2007b,kromer2009a}.  For each RT
simulation we remapped the ejecta structure to a 50${^3}$ grid. 10${^8}$
photon packets were then tracked through the ejecta for 450
logarithmically spaced time steps between 0.1 and 100 days since
explosion. We used the atomic data set described by \citet{gall2012a}. We
adopted a gray approximation in optically thick cells (cf.
\citealp{kromer2009a}) and for times earlier than 0.12 days post
explosion local thermal equilibrium (LTE) is assumed. After this, we
adopted our approximate NLTE description as presented by
\cite{kromer2009a}. We calculated line of sight dependent light curves
for 100 bins of equal solid-angle and line of sight dependent spectra
for 10 different observer orientations of equal angle spacing that lie
on the $x$-$y$-plane.  The viewing angle dependent spectra we present
here were calculated with the method described by \citet{bulla2015a}
utilizing ``virtual packets'' leading to a significant reduction in the
Monte-Carlo noise. 

\begin{figure*}[htbp]
  \centering
  \includegraphics[width=0.95\textwidth]{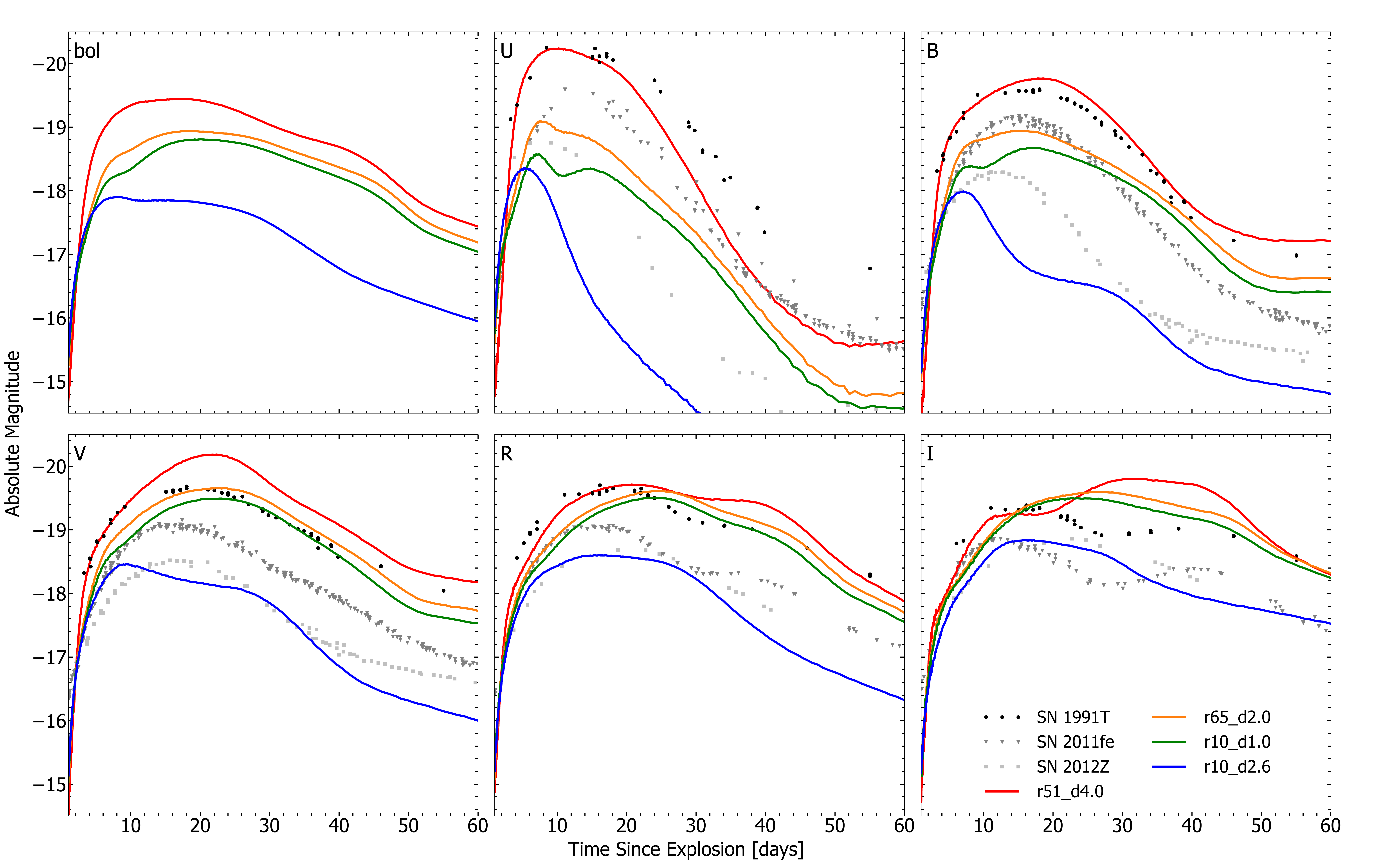}
  \caption{Angle averaged bolometric and \textit{UBVRI} band light curves for 
  a subset of our 
  PGCD models. Observed light curves representing different subclasses of 
  SNe Ia are also included.  SN 1991T (\citealp{lira1998a}), 
  SN 2011fe (\citealp{richmond2012a}, \citealp{tsvetkov2013a}, 
  \citealp{munari2013a}, \citealp{brown2014a} and 
  \citealp{stahl2019a}) and SN 2012Z  (\citealp{stritzinger2015a},
   \citealp{brown2014a} and \citealp{stahl2019a}).
   We made use of the Open Supernova Catalogue (\citealp{guillochon2017a}) 
   to obtain the observed photometry.}
  \label{fig:band_lightcurves}
\end{figure*}

Fig.~\ref{fig:band_lightcurves} shows the bolometric and \textit{UBVRI}
band angle-averaged light curves for a selection of our PGCD models
spanning from the faintest to the brightest model in our sequence.
Observed light curves representing different subclasses of SNe~Ia
relevant for comparison to our model sequence are included for
comparison: SN~1991T after which the luminous 91T-like subclass is
named, SN~2011fe, a normal SN~Ia and SN~2012Z, which is one of the
bright members of the subluminous subclass of SNe~Iax.  We have shifted
the explosion epochs of the observed light curves to better compare to
the light curve evolution of the PGCD models.  The explosion epochs we
adopted are JD 2448359.5 for SN 1991T (2 days later than estimated by
\mbox{\citealp{filippenko1992a}}), JD 2455800.2 for SN 2011fe (3 days
later than estimated by \citealp{nugent2011a}) and JD 2455955.9 for SN
2012Z (2 days later than estimated by \citealp{yamanaka2015a}).  

From Fig.~\ref{fig:band_lightcurves} we can see that a feature of our
PGCD model light curves is a relatively sharp initial rise over
approximately the first 5 days post explosion (particularly in the
\textit{U} and \textit{B} bands).  After this, the light curves become
quite flat and complex, with the exception of the faintest PGCD model
r10\_d2.6 (blue in Fig.~\ref{fig:band_lightcurves}), which shows an
initial rapid decline post peak in the \textit{U} and \textit{B} bands.
One of the intermediate brightness models shown in
Fig.~\ref{fig:band_lightcurves}, Model r10\_d1.0 (green), even exhibits
a double peak in its \textit{U} and \textit{B} band light curves.  This
peculiar light curve behavior is driven by the interplay between the
detonation ash at the center of the models and the deflagration ash
which surrounds it.  In all bands the radiation emitted from the decay
of $^{56}$Ni in the deflagration ash dominates the initial rapid rise.
However, as the band light curves approach peak, the contribution from
the radiation emitted from the deflagration ash begins to diminish while
the contribution of the radiation emitted from the detonation ash
increases to become the dominant source of the luminosity for the band
light curves. Such interplay between the deflagration and detonation ash
distinguishes these PGCD models from pure detonation models (e.g.,
\citealp{sim2010a}, \citealp{blondin2017a}, \citealp{shen2018a}) and DD
models (e.g.,
\citealp{gamezo2005a,roepke2007b,maeda2010a,seitenzahl2013a}).

It is clear from Fig.~\ref{fig:band_lightcurves} that the best match is
found between the band light curves of the brightest PGCD model in our
sequence, Model r51\_d4.0 (red) and SN~1991T. This is consistent with
the results of \cite{seitenzahl2016a} who found the best match between
their GCD model and SN~1991T.  Model r51\_d4.0 reproduces the initial
rise of SN~1991T well in all bands until approximately 10 days post
explosion with good agreement continuing to later times for certain
bands. In particular, the model shows good agreement in \textit{B} band
with SN~1991T across all epochs shown. However, the band light curves of
the model become too red as we move to later times. In particular, Model
r51\_d4.0 predicts a secondary peak that is approximately half a
magnitude brighter than the first peak.  A secondary peak of this size
in \textit{I} band is not observed for SN~1991T leading to a difference
of over half a magnitude between the model and SN~1991T around the time
of this secondary peak. 

The band light curves of our PGCD models span a wide range of peak
brightnesses (variations of almost 2 magnitudes are predicted for the
models in \textit{U}, \textit{B} and \textit{V} bands).  The models
therefore cover a range of brightnesses which encompasses luminous
SNe~Ia such as those in the 91T-like subclass, normal SNe~Ia and bright
members of the subluminous SNe~Iax class. However, the band light
curves of our PGCD models show poor agreement with the normal SNe~Ia,
SN~2011fe and the bright SNe~Iax, SN~2012Z in terms of colors and
overall light curve evolution.  Therefore, while we are able to reach
lower brightnesses with our sequence compared to previous works the
disagreement in the predicted light curves disfavors them as an
explanation for either normal SNe~Ia or SNe~Iax.

\begin{figure}[htbt]
  \centering
  \includegraphics[width=.9\linewidth,trim={0.05 1.1cm 0 0},clip]{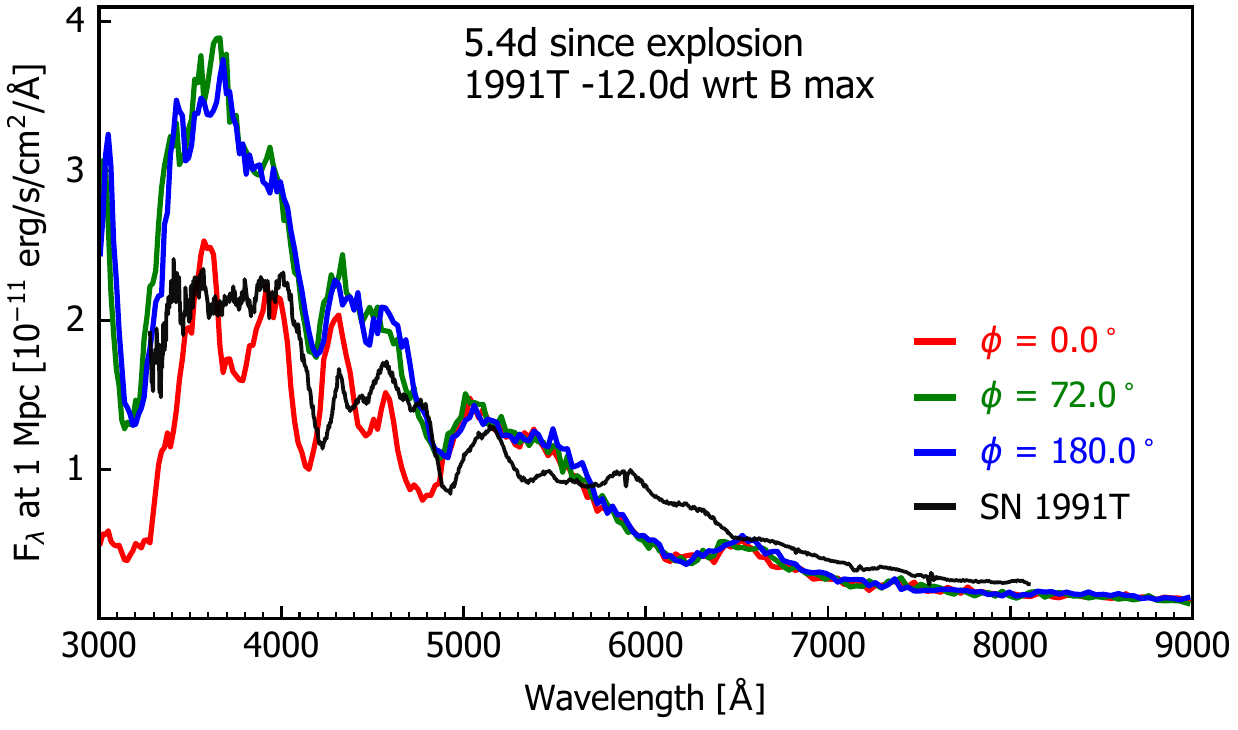}
  \includegraphics[width=.9\linewidth,trim={0.05 1.1cm 0 0},clip]{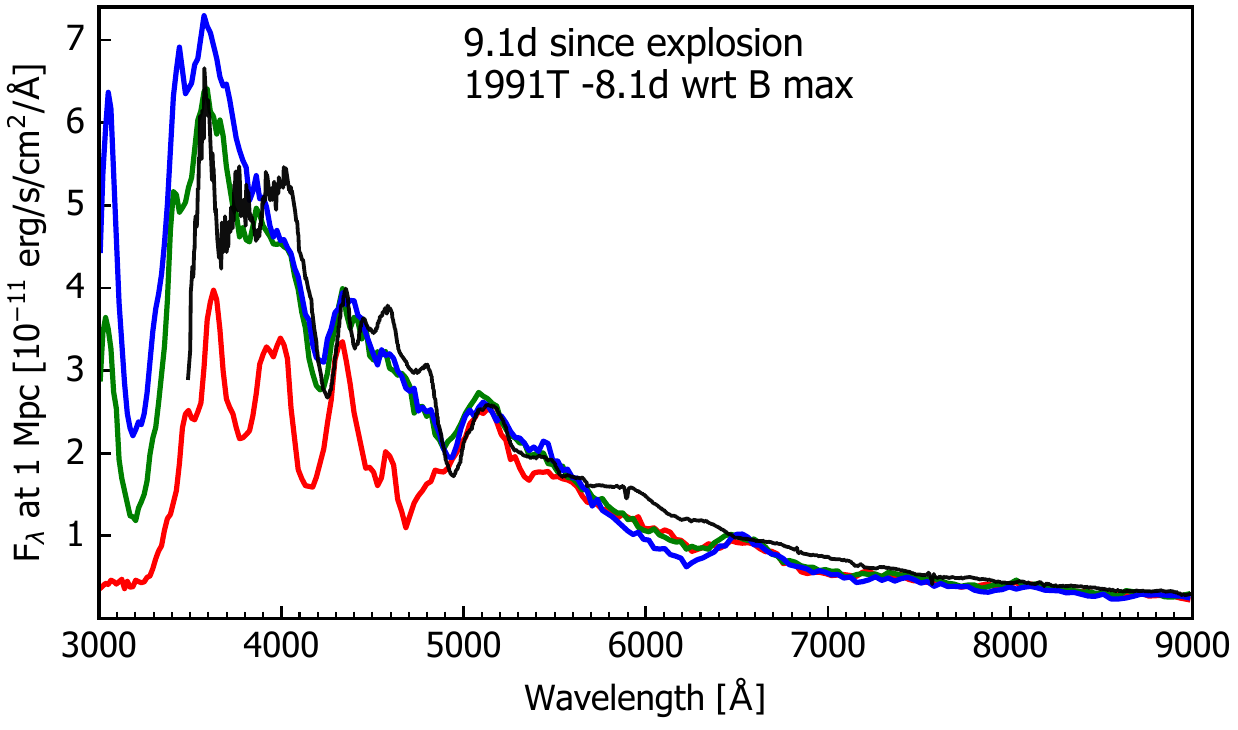}
  \includegraphics[width=.9\linewidth,trim={0.05 1.1cm 0 0},clip]{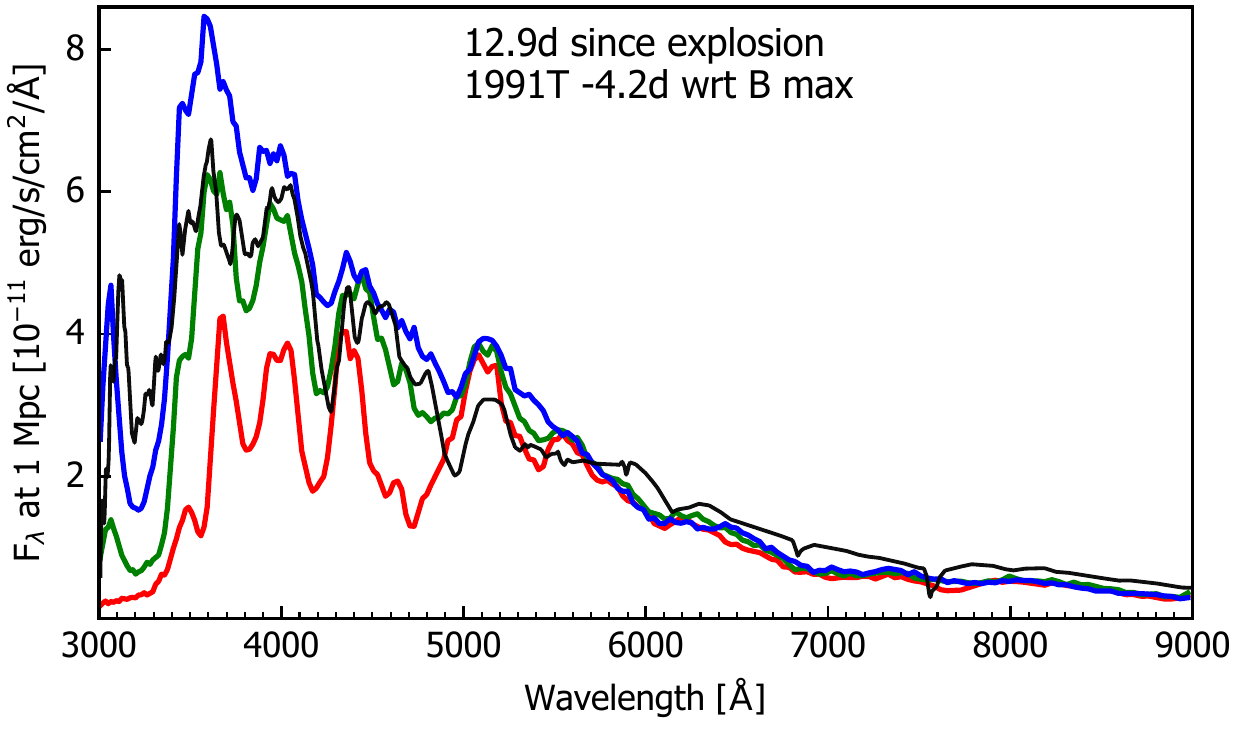}
  \includegraphics[width=.9\linewidth,trim={0.05 1.1cm 0 0},clip]{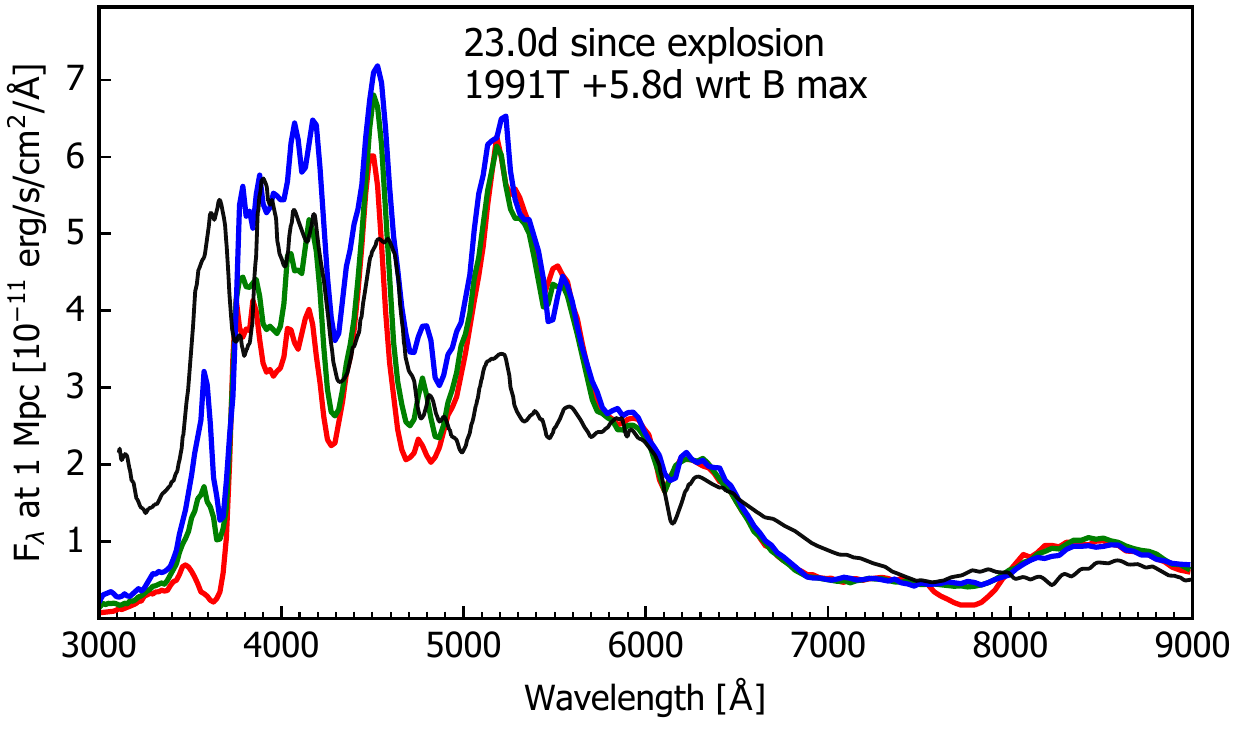}
  \includegraphics[width=.9\linewidth,trim={0.05 0.08cm 0 0},clip]{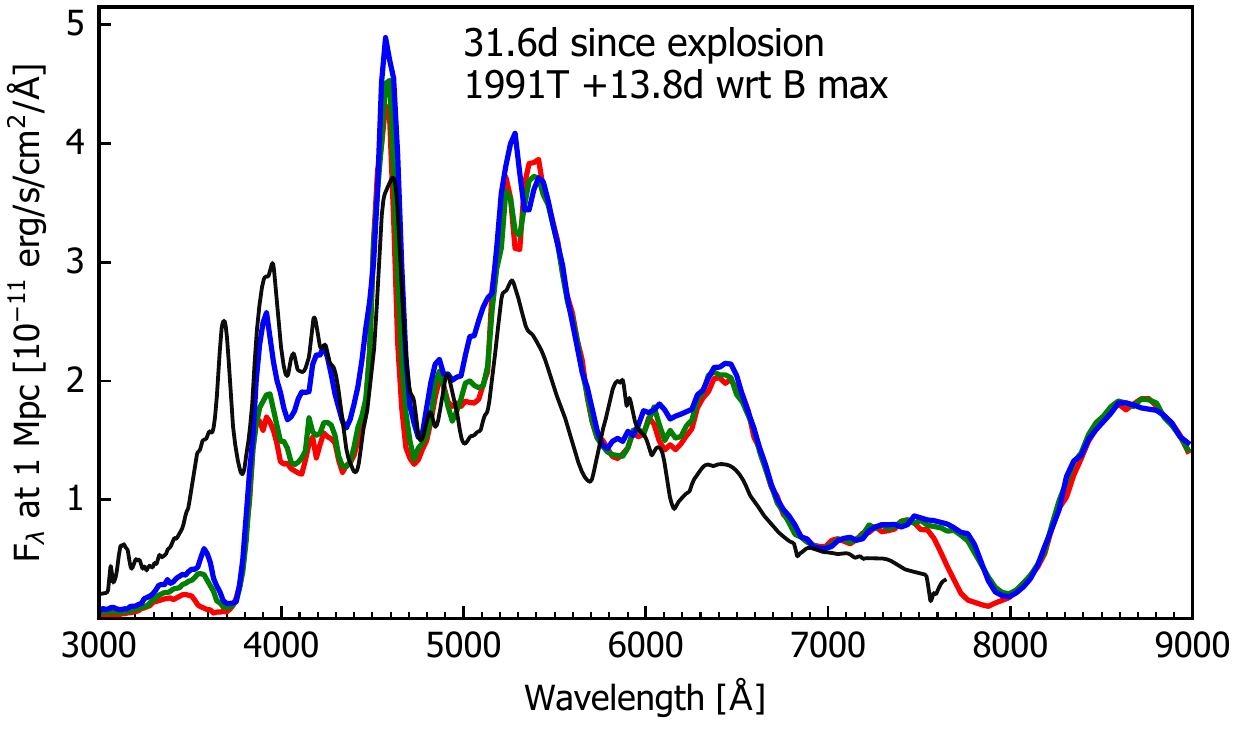}
  \caption{Spectra over a variety of epochs for a selection of viewing angles for our
  brightest PGCD model (r51\_d4.0). Observed spectra of SN 1991T 
  (\citealp{ruiz-lapuente1992a}, \citealp{phillips1992a} and \citealp{filippenko1992a}) 
  are included for comparison. The observed spectra have been 
  flux calibrated to match the photometry, dereddened using $E(B-V)=0.13$ as 
  estimated by \citealp{phillips1992a}) and deredshifted taking z = 0.006059 
  from interstellar Na.
  We take 30.76 as the distance modulus to SN 1991T (\citealp{saha2006a}).}
  \label{fig:spectra_91T}
\end{figure}

From our light curve comparisons discussed above our PGCD models appear
best suited to explain the 91T-like subclass (if any).  In
Fig.~\ref{fig:spectra_91T} we compare a selection of viewing angle
dependent spectra, over a variety of epochs, for our brightest model
(r51\_d4.0, red), this is the model which produces best overall
agreement in terms of light curves and spectra with SN~1991T. As was the
case for the GCD model of \cite{seitenzahl2016a}, our PGCD models are
almost symmetric about the axis defined by the center of the star and
the position of the initial deflagration ignition spark (located on the
positive $x$-axis in
Figs.~\ref{fig:abundslicer10d26},~\ref{fig:abundslicer60d26},
~\ref{fig:abundslicer82d1},~and~\ref{fig:abundslicer51d4}). Therefore, we
focus on different observer orientations that lie in the $x$-$y$-plane
to demonstrate the viewing angle effects here.  In
Fig.~\ref{fig:spectra_91T} we identify these with $\phi$, the angle made
between the observer orientation and the $x$-axis. The $\phi=0^{\circ}$
direction corresponds to the observer viewing from the side where the
initial deflagration spark was placed while the $\phi=180^{\circ}$
direction corresponds to the side where the detonation occurred.  The
viewing angles shown in Fig.~\ref{fig:spectra_91T} are representative of
the range of brightness exhibited by this model depending on the
line of sight.  Model spectra are shown relative to explosion as
labeled. As in Fig.~\ref{fig:band_lightcurves} we adopted an explosion
date of JD 2448359.5 for the SN~1991T spectra we compare to.

From Fig.~\ref{fig:spectra_91T} we can see that there are noticeable
viewing angle variations in the spectra of Model r51\_d4.0, for
wavelengths of $\sim$~$5000\,$\r{A} or less, at all epochs.  This
viewing angle dependency is driven by the asymmetric distribution of the
deflagration ashes in the model that are rich in IGEs. As we can see
from Fig.~\ref{fig:abundslicer51d4}, which shows the ejecta structure of
slices along the $x$-$y$-plane for Model r51\_d4.0, the deflagration
ashes extend over a greater range of velocities toward the side where
the initial deflagration was ignited at $\phi=0^{\circ}$, (positive
$x$-axis in Fig.~\ref{fig:abundslicer51d4}).  This leads to increased
line blanketing by IGEs for lines of sight looking close to this initial
deflagration spark (e.g., $\phi=0^{\circ}$ spectra, red in
Fig.~\ref{fig:spectra_91T}), and, thus, the flux in the blue and UV is
significantly reduced. The deflagration ashes, however, only cover a
significantly greater velocity range for a relatively small window of
viewing angles.  Looking along $\phi=72^{\circ}$ (green in
Fig.~\ref{fig:spectra_91T}), we already see the line blanketing by IGEs
in the blue and UV is significantly reduced and the flux at wavelengths
less than $\sim$~$5000\,$\r{A} is already quite similar to what is
observed for the viewing angle looking toward where the detonation is
first ignited ($\phi=180^{\circ}$, blue in Fig.~\ref{fig:spectra_91T}).
Wavelengths greater than $\sim$~$5000\,$\r{A} are less impacted by line
blanketing due to IGEs, and, thus, do not show significant viewing angle
dependencies.  Additionally, as we move to later times, the
viewing angle effects become less noticeable for the model as the outer
layers become more optically thin. We find the same viewing angle
dependency is observed in the \textit{B} and \textit{U} band light
curves of Model r51\_d4.0. The \textit{U} band peak magnitudes vary
between -19.5 to -20.5 while the \textit{B} band peak magnitudes vary
between -19.4 to -19.9. There is only very minimal viewing angle
variation in the \textit{VRI} band light curves. Overall, Model r51\_d4.0
exhibits viewing angle dependencies comparable to those shown by the GCD
model of \cite{seitenzahl2016a}. Indeed, we find significant
viewing angle dependencies for all of our models, with all models
showing the largest viewing angle variation in the \textit{U} and \textit{B}
bands. However, for fainter models the viewing angle variations in the
\textit{VRI} bands become increasingly prominent. We also find that the
brightest and faintest viewing angles can correspond to significantly
different observer orientations depending on the distribution of
deflagration ash and differences between the irregularly shaped core
detonations in each individual model.

The two brighter viewing angles shown in Fig.~\ref{fig:spectra_91T}
(blue and green) show best spectroscopic agreement with SN~1991T, in
particular at 9.1 and 12.9 days since explosion where they produce a
good match to the overall flux of SN~1991T across all wavelengths.
However, these viewing angles provide a poor match to the blue and UV
flux at the earliest epoch shown and also have significantly too much
flux in the red wavelengths for the last two epochs shown. Therefore,
although promising, we find no individual viewing angles which provide a
consistently good match to the spectra of SN~1991T over multiple epochs.
Additionally, as noted above and already found for the GCD model of
\cite{seitenzahl2016a}, none of the PGCD models produce good
spectroscopic agreement with normal SNe~Ia, showing IME features which
are too weak and spectra which are too red particularly toward later
epochs.

\begin{figure}[htbt]
  \centering
  \includegraphics[width=0.99\linewidth,trim={0.00 1.1cm 0 0},clip]{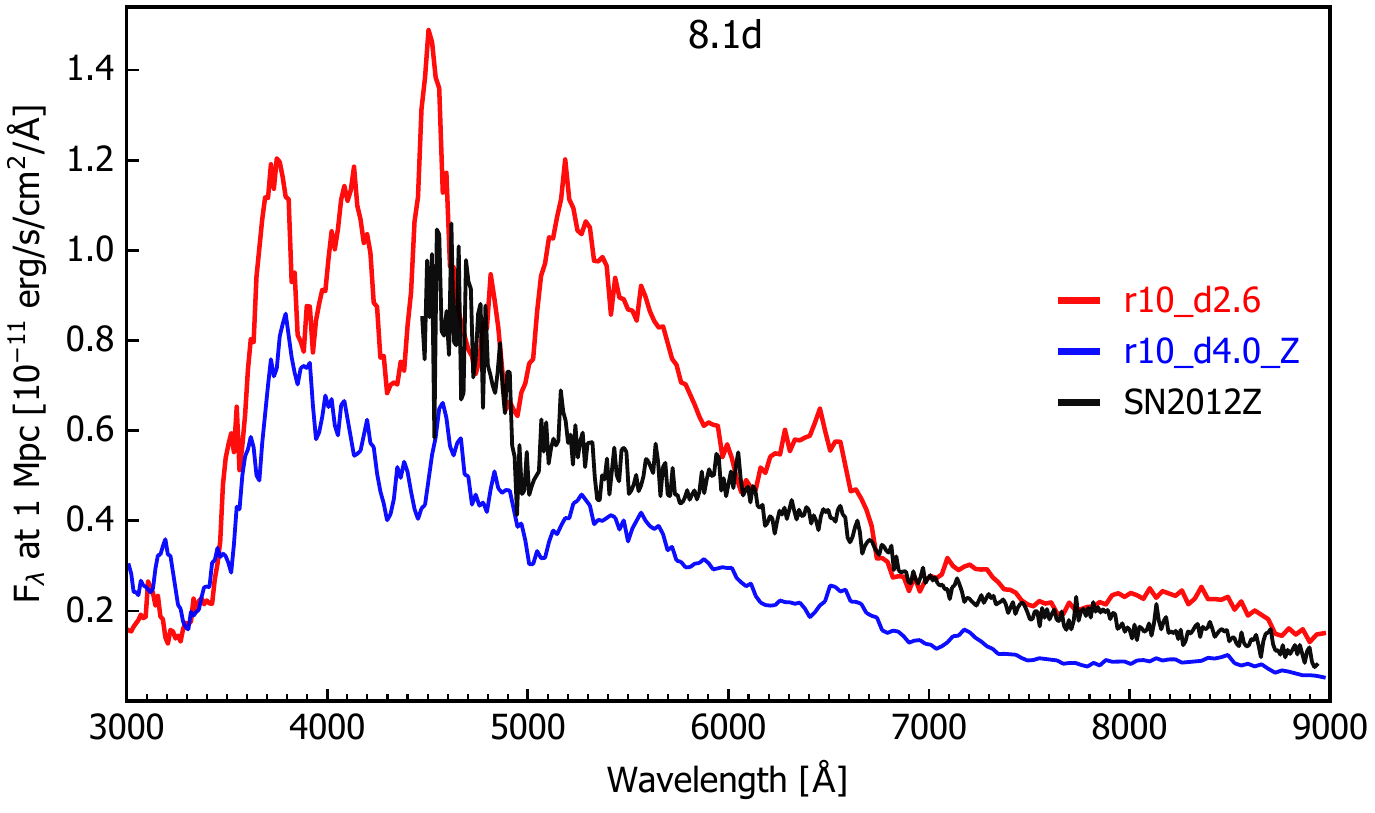}
  \includegraphics[width=0.99\linewidth,trim={0.00 0.08cm 0 0},clip]{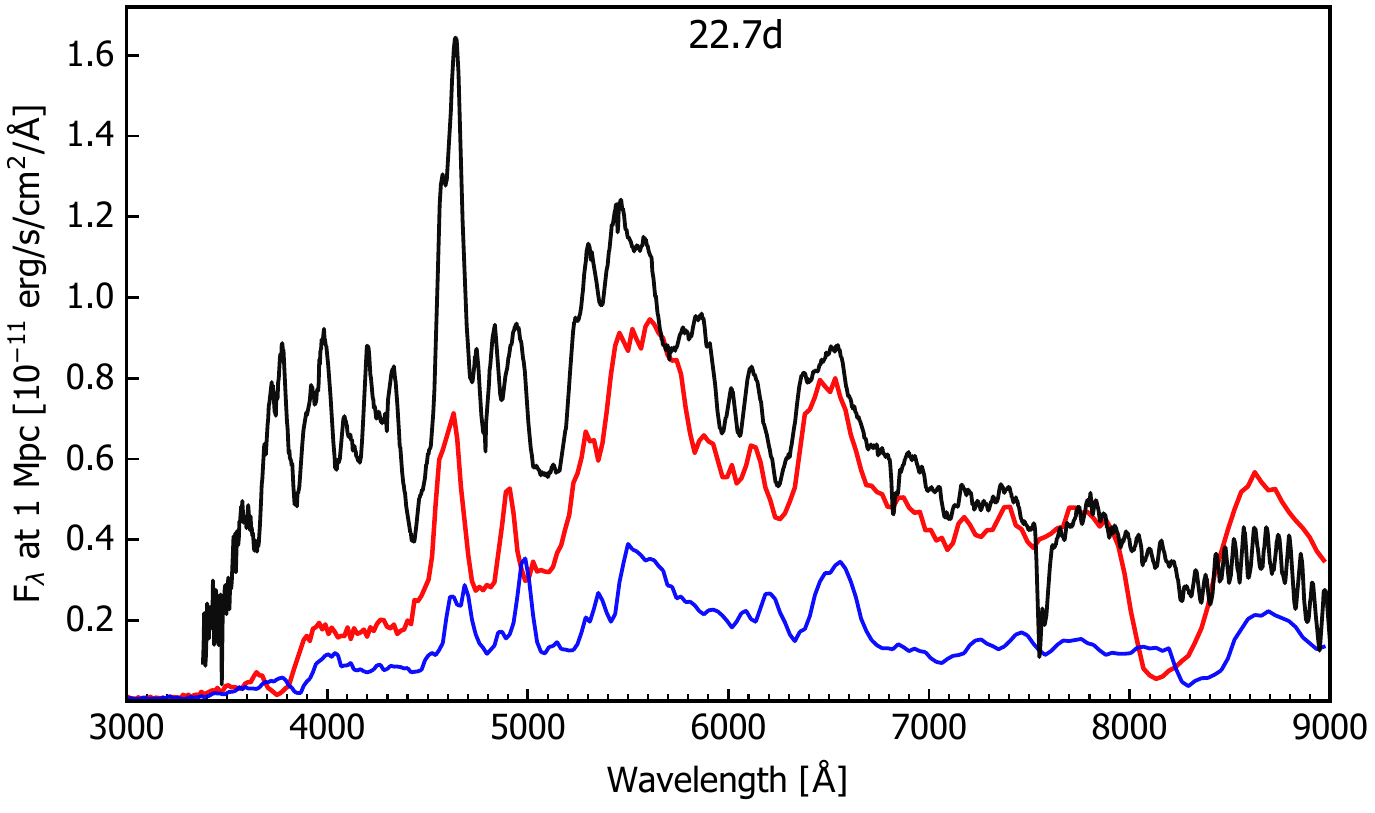}
  \caption{Spectroscopic comparisons between the bright SN~Iax, SN~2012Z, 
  (\citealp{stritzinger2015a}), along with angle-averaged spectra for
  our faintest PGCD model (r10\_d2.6, red) and the brightest pure
  deflagration model from our previous pure deflagration study, \citetalias{lach2021a}
  (r10\_d4.0\_Z, blue). Times displayed are relative to explosion for all spectra.
  We take the explosion time estimated for SN 2012Z by \cite{yamanaka2015a}.
  The observed spectra of SN 2012Z have been flux calibrated to match the 
  photometry. Additionally, they have been corrected for distance, 
  red shift and reddening assuming a distance of 33.0 Mpc, z = 0.007125 and 
  $E(B - V)=0.11$ (\citealp{stritzinger2015a}).}
  \label{fig:spectra_12Z}
\end{figure}

As we have discussed above the band light curves of the faintest PGCD
model in our sequence do not produce good agreement with those of bright
SNe~Iax such as SN~2012Z. However, our faintest model has a brightness
similar to bright SNe~Iax. As 91T-like SNe~Ia and SNe~Iax are known to
be spectroscopically similar before peak and a more stratified ejecta
structure as opposed to pure deflagration models is suggested for bright
SNe Iax \citep{barna2017a,stritzinger2015a} we briefly explore the
spectroscopic comparisons between our faintest PGCD model and the bright
SN~Iax, SN~2012Z. In particular, we investigated whether our PGCD model
spectra might compare more favorably than the pure deflagration models
of \citetalias{lach2021a}.  Fig.~\ref{fig:spectra_12Z} shows
spectroscopic comparisons between the bright SN~Iax, SN~2012Z, the
angle-averaged spectra for our faintest PGCD model (r10\_d2.6, red) and
the brightest model in our sequence of pure deflagration models from
\citetalias{lach2021a} (r10\_d4.0\_Z, blue). We compare these models to
SN 2012Z as they are the models from each sequence that match the
brightness of SN~2012Z most closely. From Fig.~\ref{fig:spectra_12Z} we
see that the spectra of our PGCD and pure deflagration models show
significantly different flux in both the epochs shown as well as a
different evolution in their flux between the two epochs. However, they
do predict many of the same spectral features, although the line
velocities of these features are noticeably more blue shifted for our
PGCD model.  This is to be expected as the PGCD model is more energetic
than the pure deflagration model leading to higher ejecta velocities.
For the earlier epoch shown (8.1 days since explosion) the pure
deflagration model (blue) shows better spectroscopic agreement with
SN~2012Z as it matches the overall flux profile of SN~2012Z much better.
At this epoch our PGCD model (red) shows a large amount of emission
between ${\sim}\, 5000\,$ to $6000\,$\r{A} as a result of the re-emission of
radiation absorbed by IGEs in the UV. This  is not observed in the
spectrum of SN~2012Z at this epoch.  For the later spectra shown (22.7
days post explosion) the PGCD model shows better overall flux agreement
with the spectra of SN~2012Z than the deflagration model. This is
because the deflagration model is too fast in decline compared to
SN~2012Z which is a general issue that is encountered when we compare
pure deflagration models to SNe~Iax (see \citetalias{lach2021a} for more
details).  However, the deflagration model reproduces the locations of
individual spectral features of SN~2012Z at least as successfully as the
PGCD model and in particular provides a better match for the location of
the Ca II infrared triplet than the PGCD model. Previous work, presented
by \citetalias{lach2021a}, has also shown that we are able to produce
reasonably good spectroscopic and light curve agreement with
intermediate luminosity SNe~Iax using our sequence of pure deflagration
models. However, none of our PGCD models presented here are faint enough
to  produce agreement with intermediate luminosity SNe~Iax.  Therefore,
our pure deflagration models do appear more promising to explain SNe~Iax
than our PGCD models. 
 
\section{Conclusions}
\label{sec:conclusion}

We used the pure deflagration simulations of \citetalias{lach2021a} as
initial models for a parameter study of the GCD scenario containing 11
models. The initial models used in this work were ignited in a single
spark off-center at varying radii $r_\mathrm{off}$ between $10$ and
$150\,\si{km}$.  Moreover, different central densities of $\rho_c=
1,\,2,\,2.6,\,3,\,4,\,5,\,6 \times 10^9\,\si{g.cm^{-3}}$ were explored.
The suite of models presented here constitutes one of the largest
systematic studies of the GCD scenario to date.

Our explosions proceed similar to the models presented in earlier works
\citep{plewa2004a,plewa2007a,townsley2007a,roepke2007a,jordan2008a,meakin2009a,seitenzahl2016a},
and \citet{byrohl2019a}: The deflagration front burns toward the
surface, burned ashes spread around the WD and collide at the opposite
side of the star. This collision produces an inwardly directed jet which
compresses and heats material ahead and leads to the initiation of a
detonation. In the case of our models, this compression is aided by the
coincidence of the first pulsation phase of the WD and the collision of
the ashes, and, thus, our models resemble the ``pulsationally assisted
gravitationally confined detonation'' scenario presented by
\citet{jordan2012b}. 

We find that all but the most energetic models from
\citetalias{lach2021a} satisfy the conditions for a detonation, that is,
they simultaneously reach values of $\rho_\mathrm{crit} > 1\times
10^7\,\si{g.cm^{-3}}$ and $T_\mathrm{crit} > 2 \times 10^9\,\si{K}$ in
one grid cell. Moreover, we can corroborate the results of
\citet{garcia2016a} who state that rotation disturbs the focus of the
collision, and, therefore, suppresses the initiation of a detonation in
the GCD mechanism. In addition, the combination of different central
densities and ignition locations introduces some diversity to the
inverse proportionality of deflagration strength and synthesized mass of
\nifs. 

The set of models produces a wide range of \nifs masses, that is,
brightnesses, from $1.057$ down to $0.257\,M_\odot$ which is the
faintest GCD model published to date. This range comprises the
subluminous SNe~Iax, normal SNe~Ia as well as the slightly overluminous
91T-like objects. We also study, inspired by \citet{flors2020a}, the
ratio of stable IGEs to \nifs in the ejecta and find that the GCD
scenario blurs the boundaries between typical nucleosynthesis products
of the sub-\mch and \mch scenario. Since our models contain a
deflagration as well as a detonation a mixture of characteristics of
both channels is reasonable. The same holds for the [Mn/Fe] value which
ranges from subsolar to supersolar values. (Detailed nucleosynthesis
yields will be published on \textsc{HESMA} \citep{kromer2017a} as will
the angle-averaged optical spectral time series and UVOIR bolometric
light curves calculated in the radiative transfer simulations.)

Furthermore, the \textit{UBVRI} band light curves show a complex
evolution reflecting the transition to a detonation at various states of
the pre-expanded WD core. We compare model light curves to the bright
SN~1991T, the normal event SN~2011fe and the SN~Iax SN~2012Z finding
that only SN~1991T can be reproduced for early times up to $10\,\si{d}$
after explosion. At later epochs ($\gtrsim 20\,\si{d}$), the light curves
exhibit an unusually prominent secondary maximum. The \textit{B} band,
however, shows good agreement with SN~1991T at all times. While there is
still some disagreement between observations and models the GCD or PGCD
scenario is still the most promising scenario for 91T-like SNe presented
to date.

We also report a strong viewing angle dependency (our analysis presented
here focuses on the brightest model in the sequence) showing line
blanketing due to the prominent deflagration ashes on the ignition side,
and, hence, a suppressed flux at blue wavelengths below ${\sim}\,
5000\,\si{\angstrom}$.  A spectral comparison to SN~1991T reveals that
good agreement is only found for intermediate times ($9.1$ and
$12.9\,\si{d}$ after explosion) for viewing angles away from the
deflagration ignition spot. For late times, however, these viewing angle
dependencies vanish since the detonation products dominate the spectra. 

Despite the poor agreement of the light curve evolution, we compared the
spectra of the faintest model in the study to SN~2012Z. In addition, we
add the brightest pure deflagration model of \citetalias{lach2021a} and
conclude that deflagrations still provide a better match to these
transients and that SNe~Iax are most probably not the result of the GCD
scenario. 

Since our simulations are not able to predict whether a detonation
is really initiated we speculate that SNe~Iax and 91T-like SNe might
originate from the same progenitor system. In the case of a failed
detonation (more likely for strong deflagrations) a SN~Iax arises and in
the case of a successful detonation (more likely for weak deflagrations)
the result is a 91T-like transient. This is a reasonable explanation for
their similar spectra near maximum light and their very different
luminosity and post-maximum evolution. It should be noted that the
ignition of a deflagration in a \mch WD is a stochastic process, and
that strong deflagrations resulting from an ignition near the center can
be considered a rare event \citep{nonaka2012a,byrohl2019a}. However,
more simmering phase simulations with different WD models are still
needed to validate the findings of \citet{nonaka2012a}. Finally,
although our fainter models do not reproduce any observations of SNe Ia,
these might still occur in nature as rare events not detected to date
and they could be identified due to their characteristic observables.

\begin{acknowledgements} 
This work was supported by the Deutsche Forschungsgemeinschaft (DFG,
German Research Foundation) -- Project-ID 138713538 -- SFB 881 (``The
Milky Way System'', subproject A10), by the ChETEC COST Action
(CA16117), and by the National Science Foundation under Grant No.
OISE-1927130 (IReNA).  FL and FKR acknowledge support by the Klaus
Tschira Foundation. FPC acknowledges an STFC studentship and SAS
acknowledges funding from STFC Grant Ref: ST/P000312/1. 

NumPy and SciPy \citep{oliphant2007a}, IPython \citep{perez2007a}, and
Matplotlib \citep{hunter2007a} were used for data processing and
plotting. The authors gratefully acknowledge the Gauss Centre for
Supercomputing e.V.  (www.gauss-centre.eu) for funding this project by
providing computing time on the GCS Supercomputer JUWELS
\citep{juwels2019} at J\"{u}lich Supercomputing Centre (JSC). Part of
this work was performed using the Cambridge Service for Data Driven
Discovery (CSD3), part of which is operated by the University of
Cambridge Research Computing on behalf of the STFC DiRAC HPC Facility
(www.dirac.ac.uk). The DiRAC component of CSD3 was funded by BEIS
capital funding via STFC capital grants ST/P002307/1 and ST/R002452/1
and STFC operations grant ST/R00689X/1. DiRAC is part of the National
e-Infrastructure.  We thank James Gillanders for assisting with the flux
calibrations of the observed spectra.
\end{acknowledgements}

\bibliography{astrofritz}
\bibliographystyle{aa}

\end{document}